\documentclass[12pt,epsfig]{article}
\usepackage{amssymb}
\usepackage[X2,T2A]{fontenc}
\usepackage[koi8-r]{inputenc}
\usepackage[english]{babel}
\usepackage{graphicx}
 
\begin{document}
\selectlanguage{english}

\thispagestyle{empty}

\bigskip
\bigskip
\bigskip
\begin{center}
\begin{Large}
{HeII$\rightarrow$HeI Recombination of Primordial Helium Plasma
Including the Effect of Neutral Hydrogen}
\end{Large}
\end{center}
\bigskip
\bigskip
\bigskip
\renewcommand{\thefootnote}{\fnsymbol{footnote}}
\begin{center}
{E. E. Kholupenko$^{1}$\footnote{eugene@astro.ioffe.ru}, 
A. V. Ivanchik$^{1,2}$, and D. A. Varshalovich$^{1,2}$}
\end{center}
\renewcommand{\thefootnote}{\arabic{footnote}}
{$^1$Ioffe Physical-Technical Institute of RAS, Russia 
\\$^2$St. Petersburg State Polytechnical University, Russia}
\bigskip
\begin{center}
\bf{Abstract}
\end{center}
{The HeII$\rightarrow$HeI recombination of primordial helium plasma 
($z = 1500 - 3000$) is considered in terms of the standard cosmological model. 
This process affects the formation of cosmic microwave
background anisotropy and spectral distortions. 
We investigate the effect of neutral hydrogen on the
HeII$\rightarrow$HeI recombination kinetics with partial and complete 
redistributions of radiation in frequency in
the HeI resonance lines. It is shown that to properly compute 
the HeII$\rightarrow$HeI recombination kinetics, one 
should take into account not only the wings in the absorption and 
emission profiles of the HeI
resonance lines, but also the mechanism of the redistribution of resonance 
photons in frequency. Thus, for
example, the relative difference in the numbers of free electrons for the 
model using Doppler absorption
and emission profiles and the model using a partial redistribution in 
frequency is 1 - 1.3\% for the epoch 
$z = 1770 - 1920$. The relative difference in the numbers of free electrons 
for the model using a partial
redistribution in frequency and the model using a complete redistribution in 
frequency is 1 - 3.8\% for the
epoch $z = 1750 - 2350$.}
\\DOI: 10.1134/S1063773708110017
\newpage
\section{Introduction}
\hspace{1.1cm}
The recombination of primordial plasma is a process
that ultimately leads to the formation of neutral
atoms from ions and free electrons due to the decrease
in temperature through cosmological expansion. This
process has three distinct epochs at which the fraction
of free electrons changes significantly: (1) HeIII$\rightarrow$HeII recombination 
($z\simeq 5000 - 7000$), 
(2) HeII$\rightarrow$HeI recombination ($z\simeq 1500 - 3000$),
and (3) HII$\rightarrow$HI recombination ($z\simeq 900 - 1600$), 
where z is the cosmological redshift. 
Since other nuclides (D, $^3$He, Li, B, etc.) in the primordial plasma 
are much fewer in number than $^1$H and $^4$He ($<10^{-4}$), 
the recombination of hydrogen-helium plasma is usually considered
(Zeldovich et al. 1968; Peebles 1968; Matsuda et al. 1969).
The recombination of other elements is considered
in isolated cases for special problems, such
as the effect of lithium recombination on the cosmic
microwave background (CMB) anisotropy (Stancil
et al. (2002) and references therein), the formation 
of primordial molecules (Galli and Palla (2002) and
references therein), etc.

The recombination of primordial plasma affects
significantly the growth of gravitational instability
and the formation of CMB spectral distortions and
anisotropy (Peebles 1965; Dubrovich 1975). The
appearance of the first experimental data on CMB
anisotropy (Relikt 1, COBE) rekindled interest in
the recombination of primordial plasma in the mid-
1980s. A number of improvements in the model
of hydrogen plasma recombination were suggested
(Jones and Wyse 1985; Grachev and Dubrovich
1991).

Significant progress in CMB anisotropy observations
achieved in the second half of the 1990s
(BOOMERANG, WMAP) necessitated including
a number of subtle effects that could affect the
recombination of primordial hydrogen and helium
at a level of 0.1 - 1\% (Leung et al. 2004; Dubrovich
and Grachev 2005; Novosyadlyj 2006; Burgin et al.
2006; Kholupenko and Ivanchik 2006; Wong and
Scott 2007; Chluba and Sunyaev 2006, 2007, 2008a,
2008b; Hirata and Switzer 2008; Sunyaev and Chluba
2008; Hirata 2008; Grachev and Dubrovich 2008).

One of the most important (for the primordial
plasma recombination kinetics) effects considered in
recent years is the absorption of HeI resonance photons
by neutral hydrogen, which leads to an acceleration
of the HeII$\rightarrow$HeI recombination (Kholupenko
et al. 2007 [hereinafter KhIV07]; Switzer and Hirata 2008; Rubino-Martin 
et al. 2007). Recently, this effect was taken into account 
by Wong et al. (2008) in the {\bf recfast} computational
code developed by Seager et al. (1999). This
code is most widely used to compute the primordial
plasma recombination kinetics when the CMB
anisotropy is analyzed. To take into account the effect
of neutral hydrogen on the HeII$\rightarrow$HeI recombination
kinetics, Wong et al. (2008) used a simple approximation
formula with adjustable parameters. The
{\bf recfast} modified in this way allows the results
of computations with the multilevel code\footnote
{By the multilevel code we mean a computational program
that uses a multilevelmodel atom.} by Switzer
and Hirata (2008) to be quickly and accurately reproduced
for any reasonable values of the cosmological
parameters. Nevertheless, the approach by Wong et al.
(2008) is inapplicable for more detailed studies of the
helium recombination, since their formula is not universal
for all resonance transitions in HeI, but can be
used only in describing the absorption of HeI $2^1P\rightarrow 1^1S$ 
resonance photons by neutral hydrogen. When
using the multilevel codes by Switzer and Hirata
(2008) and Rubino-Martin et al. (2007), which allow
the absorption of HeI $nP\rightarrow 1S$ ($n\ge 2$) resonance
photons by neutral hydrogen to be taken into account,
much of the radiative transfer calculations in the HeI
resonance lines are performed numerically, which is
computationally demanding and time-consuming.
These circumstances forced us to seek for an analytic
approach to the problem of including the effect
of neutral hydrogen on the HeII$\rightarrow$HeI recombination
kinetics that, on the one hand, would be more
universal than the approach of Wong et al. (2008) 
(i.e., would allow the effect of the absorption of HeI
$nP\rightarrow 1S$ resonance photons (where $n\ge 2$, and not
only $n=2$) on the HeII$\rightarrow$HeI recombination to be
estimated) and, on the other hand, would not reduce
the speed and accuracy of the primordial plasma recombination
computations for various parameters of
the cosmological model. This approach was implemented
on the basis of the papers by Chugai (1987)
and Grachev (1988), who analytically investigated the
diffusion of resonance radiation in the presence of
continuum absorption. 
In this paper, we present extended and more
detailed justifications of the key suggestions made
in KhIV07. We take into account
the fact that the scattering in the HeI resonance
lines (for transitions in the singlet structure of the
HeI atom) occurs with a partial redistribution in
frequency, which turns out to be important for the
results (Switzer and Hirata 2008; Rubino-Martin
et al. 2007).
Thus, our goal is to numerically compute the
HeII$\rightarrow$HeI recombination kinetics using analytic
formulas (Chugai 1987; Grachev 1988; this paper)
to allow for the peculiarities of the radiative transfer
in the HeI resonance lines (a partial redistribution
of HeI resonance photons in frequency and their
absorption in the neutral hydrogen continuum).

\section{Physical model of HeII$\rightarrow$HeI recombination} 
In the process of cosmological recombination, the
plasma deviates from its ionization equilibrium. A
necessary condition for this deviation is plasma opacity
for the intrinsic resonance recombination radiation.
This means that the emitted resonance photon is
absorbed by another neutral atom almost instantly
(compared to the characteristic recombination and
ionization time scales). Since the recombination radiation 
is excessive with respect to the equilibrium
background with a blackbody spectrum, the populations
of excited atomic states exceed their equilibrium
values. The excess of the excited-state populations
entails an increase of ionization fraction compared to
its equilibrium value and, accordingly, leads to a delay
of recombination.
In this situation, the plasma recombination can no
longer be described by the Saha formula and kinetic
equations should be invoked to describe the behavior
of the excited-state populations of neutral atoms and
the plasma ionization fraction. The total recombination
rate $J_{tot}$ [cm$^{-3}$s$^{-1}$] (dependent on $z$) 
is determined by the sum of the recombination rates (given the
forward and backward reactions) to all of the bound
HeI atomic states,
\begin{equation}
J_{tot}=\sum_{n=1}^{\infty}J_{cn}
\label{J_tot}
\end{equation}

The atomic transition rate to the ground state $J_{\rightarrow 1}$ 
is defined by the sum of the transition rates from all of
the excited states and the continuum,
\begin{equation}
J_{\rightarrow 1}=J_{c1}+\sum_{n=2}^{\infty}J_{n1}
\label{sum2}
\end{equation}
Since the number of HeI atoms in excited states at
the HeII$\rightarrow$HeI recombination epoch is not accumulated
(no more than $10^{-6}$ of the HeI atoms are in
each excited state), it may be concluded that 
$J_{tot}\simeq J_{\rightarrow 1}$. 
The quantities $J_{n1}$ in sum (2) are the differences 
between the direct, $J_{n\rightarrow 1}$, and reverse, $J_{n\leftarrow 1}$, 
transition rates,
\begin{equation}
J_{n1}=J_{n\rightarrow 1}-J_{n\leftarrow 1}
\label{J_n1}
\end{equation}

The estimates made by Zeldovich et al. (1968)
and Peebles (1968) and subsequently confirmed by
numerical calculations using multilevel model atoms
(Grachev and Dubrovich 1991; Seager et al. 2000)
showed that a simplified recombination model (the
so-called three-level model; Zeldovich et al. 1968;
Peebles 1968; Matsuda et al. 1969; Seager et al.
1999) could be used to calculate the ionization fraction
as a function of time. In this model, the recombination
rate for helium (the model energy level diagram
is presented in Fig. 1) is determined by the following
processes: the two-photon $2^1S\rightarrow 1^1S$ transitions
and the one-photon $2^1P\rightarrow 1^1S$ and $2^3P\rightarrow 1^1S$, 
i.e., three terms remain in sum (2)
and the following formula is valid: 
\begin{equation}
J_{tot}\simeq J_{ag}+J_{bg}+J_{b'g}
\end{equation}
where the subscripts denote the following states: $g\equiv 1^1S$, 
$a\equiv 2^1S$, $b\equiv 2^1P$, $b'\equiv 2^3P$ (see Fig. 1).

A proper allowance for the resonance transitions
requires a joint analysis of the kinetics of transitions
and radiative transfer in the 
$2^1P\rightarrow 1^1S$ and $2^3P\rightarrow 1^1S$ lines 
by including a number of peculiar factors, 
with the cosmological expansion and the absorption 
of HeI resonance photons by neutral hydrogen (HI) 
being the most important of them. 

According to Eq. (3), the two-photon HeI $2^1S\rightarrow 1^1S$ 
transition rate can be calculated using the formula
(for convenience, the common factor was taken
out of the brackets):
\begin{equation}
J_{ag}=A_{ag}\left(N_{a}-{g_{a}\over g_{g}}\eta^0_{ag}N_{g}\right)
\label{two_photon_rate}
\end{equation}
where $A_{ag}=51.3$ с$^{-1}$ is the coefficient of the spontaneous
two-photon $2^1S\rightarrow 1^1S$ decays, $N_{a}$ [cm$^{-3}$] is the 
population of the $2^1S$ state, $N_{g}$ [cm$^{-3}$] is the 
population of the $1^1S$ state, 
$g_{a}=1$ is the statistical weight of the $2^1S$ state, 
$g_{g}=1$ is the statistical weight of the $1^1S$ state, 
$\eta^0_{ag}$ is the equilibrium
photon occupation number at the $2^1S\rightarrow 1^1S$ ($a\rightarrow g$) 
transition frequency. 

For optically thick transitions (for HeI at the
HeII$\rightarrow$HeI recombination epoch, these are the
$n^1P\rightarrow 1^1S$ and $2^3P\rightarrow 1^1S$ 
transitions and, when the three-level model is used, the 
$b\rightarrow g$ and $b'\rightarrow g$ transitions, respectively) 
the rates of the direct and
reverse processes appearing in Eq. (3) are very close
(their relative difference can reach $10^{-9}$, depending
on n and the instant of time under consideration),
because the occupation numbers of the photon field
in HeI lines (including both equilibrium and intrinsic
HeI recombination radiations) are close to their
quasi-equilibrium values, $\eta_{fg}\simeq N_{n}/g_{n}N_{1}$. 

Thus, if $J_{n1}$ are calculated from Eq. (3), then two relatively
close numbers often has to be subtracted. In this
case, a significant loss of the computation accuracy
is possible (Burgin 2003). Therefore, the following
formula that is devoid of the above shortcoming is
used to consider the kinetics of such optically thick
transitions: 
\begin{equation}
J_{fg}=P_{fg}A_{fg}\left(N_{f}-{g_{f}\over g_{g}}\eta^0_{fg}N_{g}\right)
\end{equation}
where the subscript $f$ denotes the $nP$ state, $A_{fg}$ is the
Einstein coefficient for the spontaneous $f\rightarrow g$ transition, 
$N_{f}$ is the population of state $f$, and 
$g_{f}$ is the statistical weight of state $f$. 
The quantity $P_{fg}$ is the probability of the uncompensated 
$f\leftrightarrow g$ transitions. 

The quantity $P_{fg}$ can be calculated by jointly considering
the radiative transfer equation in the $f\rightarrow g$ line 
and the balance equation for levels $f$ and $g$. It
contains information about the effect of the intrinsic
resonance plasma radiation on the transition kinetics
with allowance made for the circumstances that accompany
the radiative transfer, such as the cosmological
expansion, the absorption of HeI resonance
photons by neutral hydrogen, etc. 

Since the $f\rightarrow g$ transition rate $J_{fg}$, which, in turn,
determines the recombination rate $J_{tot}$, directly depends
on $P_{fg}$, a proper calculation of $P_{fg}$ for the primordial
plasma conditions is one of the most important subgoals
of the cosmological recombination theory.

The methods for including $P_{fg}$ in the multilevel
code that computes the kinetic equations for the full
system of levels can be found in Seager et al. (2000),
Switzer and Hirata (2008), and Rubino-Martin et al.
(2007). The kinetic equation that describes the
HeII$\rightarrow$HeI recombination in terms of the simplified
model and the methods for including $P_{fg}$ in it can be
found in KhIV07 and Wong et al. (2008).

\section{Kinetics of HeI 2P$\rightarrow$1S resonance transitions}
Let us consider the HeI $f\rightarrow g$ resonance transition kinetics 
using HeI 2P$\rightarrow$1S (i.e. $b\rightarrow g$ and $b'\rightarrow g$), 
which mainly determine the recombination rate $J_{tot}$, as an example. 
A joint analysis of the balance equation for levels $f$ and $g$ and
the radiative transfer equation for the HeI $f\rightarrow g$ line
leads to the following formula for the probability of the
uncompensated $f\rightarrow g$ transitions:
\begin{equation}
P_{fg}=\int_{0}^{\infty}
{\kappa_{H} + \kappa_{He,f}\exp\left(-\tau_{f}\right) 
\over \kappa_{H}+\kappa_{He,f}}
\psi_{fg}(\nu)d\nu
\label{P_fg}
\end{equation}
where $\kappa_{H}$ [cm$^{-3}$s$^{-1}$Hz$^{-1}$] is the absorption coefficient
of photons ($h\nu\ge 13.6$ eV) by neutral hydrogen during ionization, 
$\kappa_{He,f}$ is the absorption coefficient
of photons by neutral helium in the $f\rightarrow g$ line, 
$\tau_{f}$ is the optical depth for the absorption of HeI resonance 
photons (including the absorption by both helium and hydrogen atoms), 
and $\psi_{fg}(\nu)$ is the emission profile in the HeI $f\rightarrow g$ line 
($\int\psi_{fg}(\nu)d\nu=1$). 

The coefficient $\kappa_{H}$ is given by the formula 
\begin{equation}
\kappa_{H}={8\pi \nu^2 \over c^2}\sigma_{H}(\nu)N_{H,1S}
\label{kappa_H}
\end{equation}
where $\sigma_{H}(\nu)$ is the photoionization cross section of
the HI ground state by a photon with frequency $\nu$, 
$N_{H,1S}$ is the number density of hydrogen atoms in the
ground state, which is equal, with a good accuracy, to
the neutral hydrogen number density $N_{HI}$.

The coefficient $\kappa_{He,f}$ ($f=b$ or $b'$) is given by the
formula:
\begin{equation}
\kappa_{He,f}={g_{f}\over g_{g}}A_{fg}N_{g}\phi_{fg}(\nu)
\label{kappa_He}
\end{equation}
where $\phi_{fg}(\nu)$ is the absorption profile in the HeI $f\rightarrow g$ 
line ($\int\phi_{fg}(\nu)d\nu=1$). The ground-state population $N_g$ 
is equal, with a good accuracy, to the total 
neutral helium number density $N_{HeI}$.

The optical depth $\tau_{f}$ is given by the formula
\begin{equation}
\tau_{f} (\nu, z)=\int_{\nu}^{\infty} {{c^3 \over 8\pi \nu'^3}
H^{-1}(z')\left(\kappa_{H}(\nu',z')+\kappa_{He,f}(\nu',z')\right)}d\nu'
\label{tau_f}
\end{equation}
where $H(z)= H_0 \sqrt{\Omega_\Lambda+\Omega_m (1+z)^3 + \Omega_{rel}(1+z)^4}$ 
is the Hubble constant (the parameters $H_0,~\Omega_\Lambda,~\Omega_m,~\Omega_{rel}$ 
are described below in Table \ref{cosm_par}), and the parameter $z'$ 
is defined by the equality $z'=(1+z)\nu'/\nu-1$.


For the convenience of the subsequent consideration,
let us represent  $P_{fg}$ as the sum of two terms: 
$P_{fg}=P^{H}_{fg}+P^{red}_{fg}$, 
where $P^{H}_{fg}$ and $P^{red}_{fg}$ are given by the formulas:
\begin{equation}
P^{H}_{fg}=\int_{0}^{\infty}
{\kappa_{H} \over \kappa_{H}+\kappa_{He,f}}
\psi_{fg}(\nu)d\nu
\label{P_H}
\end{equation}
\begin{equation}
P^{red}_{fg}=\int_{0}^{\infty}
{\kappa_{He,f}\exp\left(-\tau_{f}\right) \over \kappa_{H}+\kappa_{He,f}}
\psi_{fg}(\nu)d\nu
\label{P_r_f_expr}
\end{equation}

The quantity $P^{H}$ is the mean (averaged over the
profile $\psi_{fg}(\nu)$) destruction probability of a HeI 
$f\rightarrow g$ resonance photon as a result of its interaction with
neutral hydrogen.

The quantity $P^{red}_{fg}$ is the escape probability of  
HeI $f\rightarrow g$ photons from the line profile due to the cosmological
expansion. Note that the definition of this
quantity differs from the classical definition of the
Sobolev photon escape probability from the line profile $P^{S}_{fg}$ 
(Rybicki and dell Antonio 1993; Seager et al.
2000), being its generalization to the case that includes
the absorption of HeI resonance photons by
neutral hydrogen\footnote{Note that 
$P^{red}_{fg}$ is proportional to $(1-\bar I_{L})$, where $\bar I_{L}$  
is the integral introduced by Switzer and Hirata (2008).} 
(in this sense, $P^{red}_{fg}$ may be called
a modified photon escape probability from the line
profile).
If the absorption of HeI resonance photons
by neutral hydrogen is negligible (i.e., $\kappa_{H}/\kappa_{He,f}\ll 1$), 
then the formula for $P^{red}_{fg}$ takes the classical form:
\begin{equation}
P^{red}_{fg}|_{\kappa _{H}=0}=\int_{0}^{\infty}
\exp\left(-\tau_{He,f}(\nu)\right) \psi_{fg}(\nu)d\nu
\simeq P^{S}_{fg}\equiv \tau_{He,f}^{-1}
\left(1-\exp\left(-\tau_{He,f}\right)\right)
\label{Sobolev_probability}
\end{equation}
where $\tau_{He,f}(\nu)$ is the optical depth for photon absorption
in the HeI $f\rightarrow g$ line as a function of the frequency
and $\tau_{He,f}$ is the total optical depth for photon
absorption in the HeI $f\rightarrow g$ line given by the formula: 
\begin{equation}
\tau_{He,f}={g_{f}A_{fg}N_{g}c^3 \over g_{g}8\pi H \nu_{fg}^3}\;.
\end{equation}

Note that Switzer and Hirata (2008) and Rubino-Martin et al. (2007) 
used a different splitting into terms, namely, 
$P_{fg}=P^{S}_{fg}+\Delta P^{esc}_{fg}$, where $\Delta P^{esc}_{fg}$  
is the correction to the Sobolev escape probability due
to the presence of continuum absorption.

\subsection{The Modified Photon Escape Probability from the Line Profile}
Using (\ref{kappa_He}) and (\ref{tau_f}), we can transform Eq. 
(\ref{P_r_f_expr}) to
\begin{equation}
P^{red}_{fg}={g_{g}\over g_{f} A_{fg}N_{g}}\int_{0}^{\tau_f}
\left({\kappa_{He,f} \over \kappa_{H}+\kappa_{He,f}}\right)^2
\left({\psi_{fg}(\nu)\over \phi_{fg}(\nu)}\right)
{8\pi \nu^3 H \over c^3}\exp\left(-\tau'\right)d\tau'
\end{equation}
This integral can be roughly estimated from the formula
\begin{equation}
P^{red}_{fg}=
\left(1+\gamma^{-1}\right)^{-2}
\tau_{He,f}^{-1}
\left(1-\exp\left(-\tau_{He,f}\right)\right)
\label{P_rf_approx}
\end{equation}
where the parameter $\gamma$ is the ratio of the helium and
hydrogen absorption coefficients at the central frequency
of the $f\rightarrow g$ line and is given by the expression:
\begin{equation}
\gamma \equiv {k_{He}(\nu_{fg})\over k_{H}(\nu_{fg})}
={\left(g_{f}/g_{g}\right) A_{fg} N_{HeI}
\phi_{fg}(\nu_{fg}) c^2 \over 
\sigma_{H}\left(\nu_{fg}\right) 8\pi
\nu_{fg}^2 N_{HI}}
\end{equation}
The $z$ dependence of $\gamma$ for the $2^1P\rightarrow 1^1S$ 
and $2^3P\rightarrow 1^1S$ transitions is presented in Fig. \ref{gammas_ps}.

If the amount of neutral hydrogen is negligible (so that $\gamma \ll 1$) 
then Eq. (\ref{P_rf_approx}), as has been noted
above, transforms into the standard expression for
the Sobolev photon escape probability from the line
profile $P^{S}_{fg}$ - (\ref{Sobolev_probability}). 
If there is much neutral hydrogen (so that $\gamma \gtrsim 1$), 
then the value of $P^{red}_{fg}$ is lower than the typical Sobolev 
value of $P^{S}_{fg}$, calculated from (\ref{Sobolev_probability}). 
In KhIV07, the factor $\left(1+\gamma^{-1}\right)^{-2}$ 
in Eq. (\ref{P_rf_approx}) was discarded (i.e., in fact, 
Eq. (\ref{Sobolev_probability}))was used). For the HeI $2^1P\rightarrow 1^1S$ 
transition this neglect is valid throughout the HeII$\rightarrow$HeI 
recombination epoch, because the values of $\gamma$ for this transition are 
large (see Fig. \ref{gammas_ps}). For the HeI $2^3P\rightarrow 1^1S$ transition, 
this neglect is valid at the early HeII$\rightarrow$HeI recombination epoch 
($z\gtrsim 1900$), when the values of $\gamma$ for this transition are large 
($\gtrsim 10$, see Fig. \ref{gammas_ps}). At the late HeII$\rightarrow$HeI 
recombination epoch ($z\lesssim 1750$), when the values of $\gamma$ 
for the HeI $2^3P\rightarrow 1^1S$ transition are small ($\lesssim 1$, 
see Fig. \ref{gammas_ps}), the classical expression for the probability 
of the uncompensated transitions (\ref{Sobolev_probability}) 
is inapplicable for the HeI $2^3P\rightarrow 1^1S$ transitions. 
At the same time, however, the $2^1P\rightarrow 1^1S$ transition rate $J_{bg}$, 
which directly depends on the product $A_{bg}P^{H}_{bg}$, is so large that it
completely determines the recombination rate $J_{tot}$. 
Therefore, the accuracy of calculating $P^{red}_{b'g}$ (and, accordingly,
the $2^3P\rightarrow 1^1S$ transition rate $J_{b'g}$) does not play 
significant role.

\subsection{The Destruction Probability of a HeI Resonance
Photon during Its Interaction with Neutral Hydrogen}
The integral expression (\ref{P_H}) for $P^{H}$ can be transformed
to the following form convenient for both numerical
and approximate analytical integrations:
\begin{equation}
P^{H}_{fg}(\gamma)\simeq\int_{0}^{\infty}
{\psi_{fg}(\nu) \over 
1+\left(\phi_{fg}(\nu)/\phi_{fg}(\nu_{fg})\right)\gamma}d\nu
\label{P_H_int2}
\end{equation}
A further refinement of the form of the function $P^{H}(\gamma)$ 
depends on the approximation in which the absorption profile
$\phi_{fg}(\nu)$ is taken into account and, even
more importantly, on the specific form of the emission
profile $\psi_{fg}(\nu)$ determined by the physical conditions
under which the radiation is scattered and transferred 
in the $f\rightarrow g$ resonance line.

\subsubsection{Complete redistribution in frequency: The Doppler profile}
In the case of scattering with a complete redistribution in frequency, 
if the Doppler profile is used as the absorption, $\phi_{fg}(\nu)$, 
and emission, $\psi_{fg}(\nu)$, profiles (as was done in KhIV07)), 
the expression for $P^{H}_{fg}$ takes the form:
\begin{equation}
P^{H}_{D}(\gamma)=\int_{-\infty}^{\infty}
{\pi ^{-1/2} \exp\left(-y^2\right)\over 
1+\gamma\exp\left(-y^2\right)}dy
\label{A_H_int3}
\end{equation}
where the subscript $D$ stands for ``Doppler'', while the subscripts
$f$ and $g$ were omitted, since the function $P^{H}_{D}(\gamma)$ is 
universal for all resonance transitions when the Doppler profile is 
used. This function can be approximated
by the expression 
\begin{equation}
P^{H}_{D}(\gamma)=\left(1+p\gamma^{q}\right)^{-1}
\label{P_H_D_approx}
\end{equation}
where the parameters $p, q$ depend on the $\gamma$ range. 
Their values are given in Table \ref{tab_pq}.

\begin{table}[h]
\centering
\caption{Parameters of the approximation of $P^{H}_{D}$}
\begin{tabular}{lcc}
  \hline
  Range of $\gamma$ & p & q \\
  \hline
  $0\le \gamma \le 5\cdot 10^{2}$ & 0.66 & 0.9 \\
  $5\cdot 10^{2} < \gamma \le 5\cdot 10^{4}$ & 0.515 & 0.94 \\
  $5\cdot 10^{4} < \gamma \le 5\cdot 10^{5}$ & 0.416 & 0.96 \\
  $5\cdot 10^{5} < \gamma $ & 0.36 & 0.97 \\
  \hline
  \label{tab_pq}
\end{tabular}
\end{table}

The asymptotics of $P^{H}_{D}(\gamma)$ for $\gamma \rightarrow \infty$ 
is given by the expression (see, e.g., Ivanov 1969):
\begin{equation}
P^{H}_{D}(\gamma)\simeq {2\over \sqrt{\pi}} \gamma^{-1}\sqrt{\ln{\gamma}}
\end{equation}

The results of the calculations of the function $P^{H}_{D}(\gamma)$ 
for various $\gamma$ ranges are presented in Figs. \ref{para_P_gamma_ps} 
and \ref{orto_P_gamma_ps}. 

\subsubsection{Complete redistribution in frequency: The Voigt profile}
In the case of scattering with a complete
redistribution in frequency, if the Voigt profile is
used as the absorption, $\phi_{fg}(\nu)$ and emission, $\psi_{fg}(\nu)$ 
profiles, the probability $P^{H}$ can be approximately
calculated from the formula
\begin{equation}
P^{H}_{V}\simeq P^{H}_{D}+P^{H}_{W}
\label{approx_P_H}
\end{equation}
where the subscript $V$ stands for ``Voigt''. The quantity $P^{H}_{W}$ 
is attributable to the inclusion of the Voigt
profile wings (the subscript $W$ stands for ``wings'') 
and is given by the formula 
\begin{equation}
P^{H}_{W}=2\sqrt{a \over \gamma \pi^{3/2}}
\left({\pi \over 2} - \arctan \left(\sqrt{-a_1\sqrt{\pi} \ln{a} \over a\gamma }\right)\right)
\label{P_H_W}
\end{equation}
where $a$ is the Voigt parameter, which is defined by
the ratio of the natural (the quantum-mechanical
damping constant defining the mean level lifetime) 
and Doppler widths of level $f$: 
$a=\Gamma_{f} / 4\pi \Delta \nu_{D,f}$, $a_1\simeq 1.6$ 
is an adjustable parameter whose value is chosen 
from the condition for the best agreement between the
values of $P^{H}$ obtained from Eqs. (\ref{P_H_int2}) 
(by numerical integration) and (\ref{approx_P_H}). 
Equation (\ref{P_H_W}) can be derived 
by taking into account the fact that almost the entire
contribution to $\phi_{fg}(\nu)$ at 
$|\left(\nu-\nu_{fg}\right)/\Delta\nu_{D,f}|\gtrsim |\ln(a)|$ 
is provided by the Lorentz wings, i.e., the
following approximation is valid:
\begin{equation}
\phi_{fg}(\nu)\simeq {a\Delta\nu_{D,f} \over \pi}\left(\nu-\nu_{fg}\right)^{-2}
\label{Voigt_wings}
\end{equation}
The thermal width $\Delta \nu_{D,f}$ in Eq. (\ref{Voigt_wings}) 
is given by the expression (Lang 1978):
\begin{equation}
\Delta \nu_{D,f}={\nu_{fg}\over c}
\sqrt{{2k_BT \over m_{He}}+{2\over 3}V_{t}^2}
\label{line_width}
\end{equation}
where $m_{He}$ is the helium atomic mass, $T$ is the temperature
of the medium, and $V_{t}$ is the root-meansquare
turbulent velocity (if the distribution of turbulent
velocities is Maxwellian). In our calculations, we
assumed that $V_{t}=0$. 

The asymptotics of $P^{H}_{V}(\gamma)$ for $\gamma \rightarrow \infty$ 
is given by the expression (see, e.g., Ivanov 1969):
\begin{equation}
P^{H}_{V}\simeq \pi^{1/4}a^{1/2} \gamma^{-1/2}
\end{equation}
The results of the calculations of the function $P^{H}_{V}(\gamma)$ 
for the HeI $2^1P\rightarrow 1^1S$ and $2^3P\rightarrow 1^1S$ 
and transitions (and various $\gamma$ ranges) are presented in Figs. 
\ref{para_P_gamma_ps} and \ref{orto_P_gamma_ps} respectively. 

\subsubsection{Partial redistribution in frequency: 
The Wong-Moss-Scott approximation}
Switzer and Hirata (2008) and Rubino-Martin et al. (2007) showed
that a partial (rather than complete) redistribution in
frequency occurred at the HeII$\rightarrow$HeI recombination
epoch in the HeI $n^1P\rightarrow 1^1S$ lines. 
These authors performed a significant fraction of
their calculations (in particular, the radiative transfer
calculations) numerically. Since the problem under
consideration is complex, this is computationally
demanding and time-consuming (it takes about one
day to compute the radiative transfer in HeI lines for
one cosmological model; Rubino-Martin et al. 2007).
This approach is too resource-intensive to be used in
the three-level recombination model incorporated, in
particular, in {\bf recfast} (Seager et al. 1999; Wong et al.
2008), from which a high speed of computations (no
more than a few minutes per cosmological model) is
demanded at the required accuracy of about 0.1\%.
This necessitates seeking for an analytic solution to
the problem of radiative transfer in a resonance line in
an expanding medium in the presence of continuum
absorption or at least a suitable approximation that 
would describe satisfactorily the results of computations
with multilevel codes (Switzer and Hirata 2008;
Rubino-Martin et al. 2007). This approximation was
found by Wong et al. (2008): based on an approximation
formula of form (\ref{P_H_D_approx}), Wong et al. (2008) found that at 
$p=0.36$ and $q=0.86$ for the probability $P^{H}$ of the uncompensated 
$2^1P\rightarrow 1^1S$ transitions (let us denote it by $P^{H}_{WMS}$, 
where the subscript $WMS$ stands for ``Wong-Moss-Scott''), the simplified 
(three-level) model describes well the results of computations with
the multilevel code by Switzer and Hirata (2008) for any reasonable 
cosmological parameters. The values of $p=0.66$ and $q=0.9$ (KhIV07)
were used to determine the $2^3P\rightarrow 1^1S$ transition
probability.

The asymptotics of $P^{H}_{WMS}(\gamma)$ for $\gamma \rightarrow \infty$ 
is given by the expression:
\begin{equation} 
P^{H}_{WMS}=0.36\gamma ^{-0.86}
\label{P_H_WMS}
\end{equation}
The results of the calculations of the function $P^{H}_{WMS}$ 
are presented in Fig. \ref{para_P_gamma_ps}.

\subsubsection{Partial redistribution in frequency in the
wings: The Chugai-Grachev approximation}
The Wong-Moss-Scott approximation cannot be used to investigate 
the kinetics of HeI $n^1P\rightarrow 1^1S$ resonance transitions for $n\ge 3$, 
since the parameters $p$ and $q$ in Eq. (\ref{P_H_D_approx}) should have 
different values unique for each specific $n$ in this case. 
This necessitates
seeking for an expression for the probability of the
uncompensated transitions based on an analytic
solution of the radiative transfer equation in the HeI
resonance lines at the HeII$\rightarrow$HeI recombination
epoch.

Being formulated in full, this problem is very
complex, since it requires including a large number of
processes, such as the Hubble expansion, the partial
redistribution in frequency due to the thermal motion
of atoms, the Raman scattering, the continuum
absorption, the recoil upon scattering, etc. In this
case, the problem requires considering an integrodifferential
equation in which the redistribution function
cannot be expressed in terms of elementary
functions. Therefore, in this formulation, it is solved
mainly through computer simulations (Switzer and
Hirata 2008; Rubino-Martin et al. 2007). Nevertheless,
a number of simplifications and approximations
make it possible to reformulate the problem in a
form that allows a completely analytic solution. Thus,
for example, Chugai (1987) and Grachev (1988)
considered the diffusion of resonance line radiation in
the presence of continuum absorption. In comparison
with the complete formulation of the problem of
radiative transfer in a line, these authors disregarded
the following effects: (1) the Raman scattering was
disregarded; (2) a differential expression derived in
the diffusion approximation was used instead of the
exact integral term describing the redistribution of
photons in frequency due to the thermal motion of
atoms (see, e.g., Varshalovich and Sunyaev 1968;
Nagirner 2001), i.e., the final equation has the form
of a frequency diffusion equation for photons (Harrington
1973; Basko 1978); (3) the Doppler core of
the Voigt absorption profile was fitted by a delta function;
and (4) when formulating the mathematical
model, Grachev (1988) disregarded the expansion
of the medium, although Chugai (1987) previously
took it into account. Comparison of the papers by
Chugai (1987) and Grachev (1988) in this aspect
shows that the contributions to the probability of the
uncompensated transitions from the expansion of the
medium and the continuum absorption can be approximately
taken into account independently of each
other. Chugai (1987) derived the following formula
for the probability of the uncompensated transitions
(note that Chugai (1987) and Grachev (1988) used
notations differing from each other and from the
notation of this paper):
\begin{equation}
P^{H}_{C}=1.217a^{1/4}\gamma^{-3/4}
\label{P_H_C}
\end{equation}
where the subscript $C$ stands for ``Chugai''. 
The functional dependence in Eq. (\ref{P_H_C}) was determined from
qualitative considerations, while the numerical coefficient
was determined by numerically solving the diffusion equation.

Grachev (1988) analytically derived the formula: 
\begin{equation}
P^{H}_{G}=(8\lambda a)^{1/4}\pi^{-5/8}\gamma^{-3/4}
F\left({1\over 2}, 2, s+2, {1\over 2}\right) {\Gamma (s+3/2)\over \Gamma (s+2)}
\label{P_H_G}
\end{equation}
where the subscript $G$ stands for ``Grachev'', 
$\lambda$ is the single-scattering albedo, $F$ is the hypergeometric
function, $\Gamma$ is the gamma function\footnote
{The following approximation formulas expressible only in
terms of elementary functions can be used to calculate the
expressions containing special functions in Eq. (\ref{P_H_G}): 
\begin{equation}
{\Gamma (s+3/2)\over \Gamma (s+2)}\simeq\left(s+1.28\right)^{-1/2}
\label{Gamma_approx}
\end{equation}
\begin{equation}
F\left({1\over 2}, 2, s+2, {1\over 2}\right)\simeq
1+\exp\left(-1.07\ln\left(s+1.5\right)-0.45\right)
\label{F_approx}
\end{equation}
These formulas are valid in the interval of $s$ $[-1/4;\infty)$ with a
relative accuracy of at least 0.5\%.
}, 
and the parameter $s$ is defined by the formula:
\begin{equation}
s=2^{-3/2}\pi^{-1/4}(1-\lambda)\lambda^{-1/2}a^{1/2}\gamma^{1/2}-1/4
\end{equation}
Equation (\ref{P_H_G}) has the following asymptotics:
\\1) at small $\gamma$ 
($\gamma \ll \gamma_1\equiv \pi^{1/2} \lambda \left(2a(1-\lambda)^2\right)^{-1}$):
\begin{equation}
P^{H}_{G}\simeq 1.217 (\lambda a)^{1/4}\gamma^{-3/4}
\label{P_H_G_low}
\end{equation}
\\2) at large $\gamma$ ($\gamma \gg \gamma_1$)
\begin{equation}
P^{H}_{G}\simeq 2^{3/2}\pi^{-1/2}(1-\lambda)^{-1/2}\lambda^{1/2}\gamma^{-1}
\label{P_H_G_high}
\end{equation}

We see that at small $\gamma$ ($\gamma \ll \gamma_1$, 
which corresponds to $\gamma \ll 5\cdot 10^8$ for the 
HeI $2^1P\rightarrow 1^1S$ transition) and $\lambda$ close to unity, Eq. 
(\ref{P_H_G_low}) matches the solution obtained by Chugai (1987).

It should be noted that Eqs. (\ref{P_H_C}) and (\ref{P_H_G}) were
derived by Chugai (1987) and Grachev (1988) for $\gamma$ at
which $a\gamma \gg 1$ and cannot be applied at $a\gamma \lesssim 1$. 
This can be seen from the following: 
1) when $\gamma \rightarrow 0$, $P^{H}_{C}$ and $P^{H}_{G}$ tend to infinity; 
2) when the Voigt parameter tends to zero ($a\rightarrow 0$), $P^{H}_{C}$ and 
$P^{H}_{G}$ also tend to zero, while the probability $P^{H}$ that they 
describe tends to $P^{H}_{D}$ that corresponds to the correct description 
of the Doppler core (in terms of the Doppler profile rather than the delta 
function, as was done by Chugai (1987) and Grachev (1988)). 
It should also be noted that
the correct description of the Doppler core leads to
a correction of about 10\% to $P^{H}_{G}$ for the HeI $2^1P\rightarrow 1^1S$ 
transition in the range $10^6 \le \gamma \le 10^9$. 

The results of the calculations of the function $P^{H}_{G}$, its asymptotics 
(\ref{P_H_G_low}) and (\ref{P_H_G_high}), and the relative error
in $P^{H}_{G}$ calculated from the approximation formulas 
(\ref{Gamma_approx}) and (\ref{F_approx}) compared to the calculations based
on the exact formula (\ref{P_H_G}) are presented in Fig. \ref{P_CG_ps}. 
Comparison of asymptotics (\ref{P_H_G_low}) and (\ref{P_H_G_high}) 
(Fig. \ref{P_CG_ps}) shows that the $\gamma$ dependence of $P^{H}_{G}$, 
changes significantly at $\gamma$ close to $\gamma_1$ 
(e.g., on a logarithmic scale, the slope 
$\partial \ln {P^{H}_{G}}/\partial \ln {\gamma}$ changes from -0.75 to -1). 
This should be taken into account when the
Chugai-Grachev approximation is used to calculate
the HeII$\rightarrow$HeI recombination kinetics, because 
$\gamma$ (for the HeI $2^1P\rightarrow 1^1S$ transition)
takes on values close to $\gamma_1\sim 5\cdot 10^8$ 
during the HeII$\rightarrow$HeI recombination
(at the epochs $z=1900 - 2100$) (see Fig. \ref{gammas_ps}).

\subsubsection{Partial redistribution in frequency: An approximate
allowance for the Raman scattering}
Since Chugai (1987) and Grachev (1988) (1) disregarded
the Raman scattering and (2) described the
central region of the absorption profile by $\delta$-function, 
Eq. (\ref{P_H_G}) gives an underestimated value compared to
the probability $P^{H}$, determined when solving the full
problem, in which the Raman scattering contributing
to the formation of broader wings in the emission profile 
$\psi_{fg}(\nu)$, is taken into account and the central region
of the absorption profile (Doppler core) is described
by the Voigt profile. To estimate the contribution from
the Raman scattering and the Doppler core to the
probability of the uncompensated transitions $P^{H}$, 
let us represent Eq. (\ref{P_H_int2}) as:
\begin{equation}
P^{H}\simeq P^{H}_{D}+P^{H}_{G}+P^{H}_{R}
\label{P_H_int3}
\end{equation}
where $P^{H}_{D}$, $P^{H}_{G}$ and $P^{H}_{R}$ are the contributions to the
probability $P^{H}$, from various physical effects. These
quantities are explained in detail below. 

1) The contribution $P^{H}_{D}$ can be defined by the formula:
\begin{equation}
P^{H}_{D}=\int_{A}{\psi_{fg}(\nu) \over 
1+\left(\phi_{fg}(\nu)/\phi_{fg}(\nu_{fg})\right)\gamma}d\nu\;,
~~~~~~~~A\simeq [\nu_{fg}-3\Delta\nu_{D}; \nu_{fg}+3\Delta \nu_{D}]
\end{equation}
where $A$ is the Doppler core region. The quantity $P^{H}_{D}$ 
includes the contribution from the absorption of
HeI resonance photons from the Doppler core region
to the probability of the uncompensated HeI $2^1P\rightarrow 1^1S$, 
which is disregarded in the Chugai-Grachev approach. The quantity 
$P^{H}_{D}$  can be calculated using Eq. (\ref{P_H_D_approx}).

2) The contribution $P^{H}_{G}$ can be defined by the formula:
\begin{equation}
P^{H}_{G}=\int_{B}{\psi_{fg}(\nu) \over 
1+\left(\phi_{fg}(\nu)/\phi_{fg}(\nu_{fg})\right)\gamma}d\nu\;,
~~~~~~~~B\simeq [\nu_{fg}-\nu_{CG}; \nu_{fg}-3\Delta \nu_{D}] 
\cup [\nu_{fg}+3\Delta\nu_{D}; \nu_{fg}+\nu_{CG}]
\end{equation}
The quantity $P^{H}_{G}$ includes the contribution from the
region of the ``near'' wings $B$, where the redistribution
in frequency due to the thermal motion of atoms plays a crucial 
role in forming the emission profile, just as in the Doppler core region.
$P^{H}_{G}$ can be calculated using Eq. (\ref{P_H_G}). 
The characteristic frequency $\nu_{CG}$ specifies the boundaries of 
the frequency region $B$ in such a way that the redistribution in
frequency due to the thermal motion of atoms considered
by Chugai (1987) and Grachev (1988) has a
decisive effect on the formation of the emission profile
in the frequency range $[\nu_{fg}-\nu_{CG}; \nu_{fg}+\nu_{CG}]$, 
while the Raman scattering has a decisive effect on the formation
of the emission profile in the frequency ranges 
$[0; \nu_{fg}-\nu_{CG}]$ and $[\nu_{fg}+\nu_{CG}; \infty]$.  
We calculate $\nu_{CG}$ below.

3) The contribution $P^{H}_{R}$ can be defined by the formula:
\begin{equation}
P^{H}_{R}=\int_{C}{\psi_{fg}(\nu) \over 
1+\left(\phi_{fg}(\nu)/\phi_{fg}(\nu_{fg})\right)\gamma}d\nu\;,
~~~~~~~~C\simeq [0; \nu_{fg}-\nu_{CG}] \cup [\nu_{fg}+\nu_{CG}; \infty]
\label{P_H_R}
\end{equation}
The quantity $P^{H}_{R}$ includes the contribution from the
region of the ``far'' wings $C$, where the Raman scattering
(the subscript $R$ stands for ``Raman'') plays a crucial role 
in forming the emission profile.

Since the emission and true absorption\footnote
{The term ``true absorption'' is used here in the same sense as
that in Ivanov (1969) and Nagirner (2001).}
profiles in the ranges of integration in (\ref{P_H_R}) are defined by 
the expressions $\psi_{fg}(\nu)=(1-\lambda)\phi_{V}(\nu)$, 
and $\phi_{fg}(\nu)=(1-\lambda)\phi_{V}(\nu)$ (where $\phi_{V}(\nu)$ is 
the Voigt profile), we obtain
\begin{equation}
P^{H}_{R}=2(1-\lambda)\sqrt{a \over (1-\lambda)\gamma \pi^{3/2}}
\left({\pi \over 2} - 
\arctan \left(f(a, \lambda, \gamma)\right)\right)
\label{P_H_R2}
\end{equation}
where $f(a, \lambda, \gamma)$ is defined by the expression
\begin{equation}
f(a, \lambda, \gamma)=A_{1}\pi^{1/4}\left((1-\lambda)a\gamma\right)^{-1/2}
\left({\nu_{CG}\over \Delta \nu_D}\right)
\label{f_limit}
\end{equation}
Here, $A_{1}$ is the correction factor that takes into account
the fact that the two effects (both the redistribution
in frequency due to the thermal motion of
atoms and the Raman scattering) give comparable
contributions to the probability $P^{H}$ at frequencies
close to the boundary frequency $\nu_{fg}-\nu_{CG}$ (and, accordingly, 
$\nu_{fg}+\nu_{CG}$) (i.e., there is no well-defined
boundary between the zones of influence of these
effects).

Using the results by Chugai (1987) and Grachev (1988), 
one can show that the integrand in (\ref{P_H_int3}) 
in the central frequency region $[\nu_{fg}-\nu_{CG}; \nu_{fg}+\nu_{CG}]$ 
depends on the frequency as 
\begin{equation}
{\psi_{fg}(\nu) \over 
1+\left(\phi_{fg}(\nu)/\phi_{fg}(\nu_{fg})\right)\gamma}\sim
\exp\left(-{\left(x \over 2^{1/4}\pi^{-1/8}(\lambda a \gamma)^{1/4}
\right)^2}\right)
\label{integr_CG}
\end{equation}
where $x=(\nu-\nu_{fg})/\Delta \nu_{D}$.

In the frequency ranges $[0; \nu_{fg}-\nu_{CG}]$ and $[\nu_{fg}+\nu_{CG}; \infty]$, 
where the effect of Raman scattering prevails,
the integrand in (\ref{P_H_int3}) depends on the frequency as
\begin{equation}
{\psi_{fg}(\nu) \over 
1+\left(\phi_{fg}(\nu)/\phi_{fg}(\nu_{fg})\right)\gamma}\sim
{a (1-\lambda) \over \pi x^2 +a(1-\lambda)\gamma \sqrt{\pi}}
\label{integr_our}
\end{equation}
Comparing Eqs. (\ref{integr_CG}) and (\ref{integr_our}), we can 
find the characteristic frequency $\nu_{CG}$ in the form 
\begin{equation}
\nu_{CG}\simeq 2^{1/4}\pi^{-1/8}(\lambda a \gamma)^{1/4}
\sqrt{\ln{\gamma}}\Delta\nu_{D}
\label{nu_CG}
\end{equation}
Substituting (\ref{nu_CG}) into (\ref{f_limit}), we obtain the final expression  
for $f(a, \lambda, \gamma)$:
\begin{equation}
f(a, \lambda, \gamma)\simeq A_{1}2^{1/4}\pi^{1/8}
(1-\lambda)^{-1/2}\lambda^{1/4}\left( a \gamma\right)^{-1/4}\sqrt{\ln{\gamma}}
\label{f_limit_final}
\end{equation}

The coefficient $A_1=0.5$ can be determined from
the condition for the best agreement between the
dependences $P_{bg}(z)=(P^{red}_{bg}(z)+P^{H}_{bg}(z))$ 
(see Fig. \ref{para_P_z_ps}), calculated here and in 
Rubino-Martin et al. (2007).

It should be noted that we assume here that if
the fraction of coherent scatterings (i.e., the singlescattering
albedo) is equal to $\lambda$, then the fraction of
Raman scatterings is equal to $(1-\lambda)$, i.e., only the
radiative transitions are taken into account, while the
effect of the transitions produced by electron collisions
is considered negligible. This approximation is
valid at the HeII$\rightarrow$HeI recombination epoch. In the
general case where the effect of electron collisions
is not negligible, the fraction of Raman scatterings
is not equal to $(1-\lambda)$. This should be taken into
account when the above formulas are used. 

As the fraction of coherent scatterings $\lambda$ changes,
Eq. (\ref{P_H_int3}) has the following asymptotics:
\\1) As the fraction of coherent scatterings $\lambda$ tends
to unity, Eq. (\ref{P_H_int3}) turns into the sum of $P^{H}_{D}$ and 
$P^{H}_{C}$ given by Eq. (\ref{P_H_C}) (Chugai 1987). 
This limit corresponds to the absence of Raman scatterings, i.e.,
the resonance photons are redistributed in frequency
solely through the thermal motion of atoms.
\\2) As the fraction of coherent scatterings $\lambda$ tends
to zero, Eq. (\ref{P_H_int3}) tends to $P^{H}_{V}$ given by 
Eq. (\ref{approx_P_H}). This limit corresponds to a completely incoherent
scattering, which leads to a complete redistribution of
resonance photons in frequency. As a result, the Voigt
emission profile is formed.

\section{The Cosmological Model}
We performed all calculations within the framework
of standard cosmological $\Lambda$CDM models. 
The calculation results presented in Figs. \ref{var_model_ps}, 
\ref{para_P_z_ps} and \ref{orto_P_z_ps} were obtained 
using the cosmological parameters from Rubino-Martin et al. (2007) 
(since these figures reflect, in particular, the comparison of our results
with those of Rubino-Martin et al. (2007)).
The results of calculations presented in Fig. \ref{omega_bar_var_ps} 
were obtained using the cosmological parameters from Table \ref{cosm_par}. 

\begin{table}
\centering
\caption{Parameters of the standard cosmological model}
\begin{tabular}{lll}
  \hline
  Description & Designation & Value \\
  \hline
  Total matter density & $\Omega_{tot}$ & ~~~~1 \\
  (in units of critical density) & & \\
  Baryonic matter density& $\Omega_b$ & $0.02 - 0.06$ \\
  Nonrelativistic matter density& $\Omega_m=\Omega_{CDM}+\Omega_{b}$ & $0.27$ \\
  Relativistic matter density & $\Omega_{rel}=\Omega_{\gamma}+\Omega_{\nu}$ & $8.23\cdot 10^{-5}$ \\
  Vacuum-like matter density& $\Omega_\Lambda$ & $0.73$\\
  Hubble constant & $H_0$ & 70 km/s/Mpc \\
  CMBR temperature today & $T_{0}$ & $2.726$ K \\
  Helium mass fraction & $Y$ & $0.24$ \\
  \hline
  \label{cosm_par}
\end{tabular}
\end{table}

\section{Results and Discussion}
The calculated destruction probabilities of resonance
photons as they interact with neutral hydrogen $P^{H}$ 
are plotted against the ratio of the helium and
hydrogen absorption coefficients $\gamma$ in Figs. 
\ref{para_P_gamma_ps} and \ref{orto_P_gamma_ps} for
various models of photon redistribution in frequency.
As we see from these figures, the probabilities $P^{H}$ 
are lowest and highest when the Doppler and Voigt
profiles, respectively, are used to describe the absorption
and emission coefficients. This is because the Doppler profile 
``has no wings'', while the Voigt profile has slowly descending 
Lorentzian wings. Therefore, the calculated $P^{H}$ 
for any other models of redistribution
(for a partial redistribution) in frequency (including
the actual dependence $P^{H}(\gamma)$) should lie between
the curves corresponding to the use of the Doppler
and Voigt profiles (i.e., the inequality $P^{H}_{D}\le P^{H} \le P^{H}_{V}$ 
should hold). 

As was shown by Switzer and Hirata (2008) and
Rubino-Martin et al. (2007), the partial redistribution
in frequency should be taken into account when the
probability $P^{H}$ is calculated for the HeI $2^1P\rightarrow 1^1S$ 
transition, because the fraction of coherent scatterings
in the total number of resonance photon scatterings
in the line is high, $\lambda\simeq (1-2.5\cdot 10^{-3})$. The
probability of the uncompensated HeI $2^1P\rightarrow 1^1S$ 
transitions can be calculated by taking into account
the partial redistribution in frequency using Eq. (\ref{P_H_int3}) 
or in the Wong-Moss-Scott approximation (\ref{P_H_WMS}) 
(Fig. \ref{para_P_gamma_ps}). In the range of values $10^7\le \gamma \le 10^9$
important for the HeII$\rightarrow$HeI recombination kinetics, the
calculated values of $P^{H}$ and $P^{H}_{WMS}$ are in satisfactory
agreement (differ by no more than 50\%, although the
values of the quantities themselves change by two
orders of magnitude).

The scattering in the HeI $2^3P\rightarrow 1^1S$ resonance
line occurs at an almost complete redistribution of
photons in frequency due to the high fraction of incoherent
(Raman) scatterings,$(1-\lambda)\simeq 1$ (and, accordingly, 
the low fraction of coherent scatterings, $\lambda\simeq 0$). 
Therefore, Eq. (\ref{approx_P_H}) which was derived by assuming 
a complete redistribution of resonance photons 
in frequency, is used to calculate the probability $P^{H}$ 
(equal to $P^{H}_{V}$) for this transition.
Figure \ref{orto_P_gamma_ps} presents the calculated values of 
$P^{H}_{V}$ and $P^{H}_{D}$ for the HeI $2^3P\rightarrow 1^1S$ transition. 
We see from Fig. \ref{orto_P_gamma_ps} that the curves for $P^{H}_{D}$ and 
$P^{H}_{V}$ coincide in the range of values $10^{-2}\le \gamma\le 10^4$ 
important for the transition under consideration at the HeII$\rightarrow$HeI 
recombination epoch (see Fig. \ref{gammas_ps}). 
This is because the contribution from the wings $P^{H}_{W}$ to the probability 
$P^{H}_{V}$ at these values of $\gamma$ is negligible compared to the 
contribution from the Doppler core $P^{H}_{D}$. In turn, this is because
the Voigt parameter is small for the HeI $2^3P\rightarrow 1^1S$ transition: 
$a\simeq 10^{-5}$. The contribution from the wings $P^{H}_{W}$ 
to the probability $P^{H}$ begins to have an affect at
such values of $\gamma$ that $a\gamma\gtrsim 1$ (this is true not only for
a complete redistribution in frequency, but also for a partial redistribution 
in frequency). Thus, in HeII$\rightarrow$HeI recombination calculations, 
the probability $P^{H}$ for the HeI $2^3P\rightarrow 1^1S$ transition can 
be calculated with a sufficient accuracy from Eq. (\ref{P_H_D_approx}) 
(as was done in KhIV07), although it would be more appropriate to calculate 
this probability from Eq. (\ref{approx_P_H}). 

The main result of this paper is the dependence of the relative 
number of free electrons $N_{e}/N_{H}$\footnote
{$N_{H}$ is the total number of hydrogen atoms and ions} 
on redshift $z$ for the HeII$\rightarrow$HeI recombination epoch 
(Fig. \ref{var_model_ps}), calculated by including the effect of 
neutral hydrogen when using various models for the redistribution of
HeI resonance photons in frequency upon their scattering
in the HeI $2^1P\rightarrow 1^1S$ line. 
Figure \ref{var_model_ps} leads us to the following conclusions: 
1) The effect of neutral hydrogen on the HeII$\rightarrow$HeI recombination
turns out to be significant, since including it changes $N_{e}/N_{H}$ 
by $1 - 2.1\%$ for the epochs $z=1650 - 1950$ compared to the recombination 
scenario in which it is disregarded (Dubrovich and Grachev 2005; Wong
and Scott 2007).
This change in $N_{e}/N_{H}$ is significant
for a proper analysis of the experimental data
on CMB anisotropy that will be obtained from the
Planck experiment scheduled for 2009; 
2) The calculated $z$ dependence of $N_{e}/N_{H}$ at the HeII$\rightarrow$HeI
recombination epoch turns out to be sensitive to the model for redistribution 
of HeI resonance photons in frequency. 
Thus, for example, the relative difference in 
$N_{e}/N_{H}$ for the model using the Doppler absorption
and emission profiles (KhIV07) and
the model using a partial redistribution in frequency
(Switzer and Hirata 2008; Rubino-Martin et al. 2007;
Wong et al. 2008; this paper) is $1 - 1.3\%$ for the epochs 
$z=1770 - 1920$. The relative difference in $N_{e}/N_{H}$ 
for the model using a partial redistribution in frequency
and the model using a complete redistribution in frequency
for the HeI $2^1P\rightarrow 1^1S$ resonance transition
(Rubino-Martin et al. 2007; this paper) is $1 - 3.8\%$ 
for the epochs $z=1750 - 2350$. 
These differences are
significant for the analysis of the experimental data
from CMB anisotropy measurements during future
experiments (Planck etc.). This suggests that simple
models for the redistribution of resonance photons in
frequency (such as those using the Doppler emission
profile or a complete redistribution in frequency) are
inapplicable for describing the effect of neutral hydrogen
on the HeII$\rightarrow$HeI recombination kinetics
and that the partial redistribution should be taken
into account by using either numerical calculations
(Switzer and Hirata 2008), or fitting formulas (Wong
et al. 2008), or an analytic solution (this paper), or
a combination of these approaches (Rubino-Martin
et al. 2007). 

Our calculated values of $N_e/N_H$ agree with those
calculated by Rubino-Martin et al. (2007) (for the
model with a partial frequency redistribution of the 
resonance photons produced during transitions in the
HeI singlet structure) with a relative accuracy of
at least $10^{-3}$ (see Fig. \ref{var_model_ps}, bottom panel). 
Our values of $N_e/N_H$ calculated in the approximation of 
a complete redistribution of HeI resonance photons in frequency
(i.e., using Eq. (\ref{approx_P_H}) for both $2^3P\rightarrow 1^1S$ and 
$2^1P\rightarrow 1^1S$ lines) agree with those of Rubino-Martin et al. (2007) 
(for a complete redistribution in frequency in the $2^3P\rightarrow 1^1S$ 
and $2^1P\rightarrow 1^1S$ lines) with a relative accuracy of at least 
$4\cdot 10^{-3}$ (see Fig. \ref{var_model_ps}, bottom panel). 

Our additional results to be compared with those
of other authors are the dependences of the probabilities 
$P_{fg}=(P^{H}_{fg}+P^{red}_{fg})$ for the HeI $2^1P\rightarrow 1^1S$ 
(Fig. \ref{para_P_z_ps}) and $2^3P\rightarrow 1^1S$ (Fig. \ref{orto_P_z_ps}) 
transitions on redshift $z$ for various models of the redistribution of HeI
resonance photons in frequency upon their scattering
in lines. The calculated values of these quantities are
in satisfactory agreement with those of Switzer and
Hirata (2008) and Rubino-Martin et al. (2007). 

In conclusion, we calculated $N_{e}/N_{H}$ for various 
values of the cosmological parameters. It should be 
noted that the $z$ dependence of $N_{e}/N_{H}$ changes only 
slightly when varying the fraction of nonrelativistic 
matter $\Omega_m$ within the range $0.24 - 0.30$ and the Hubble
constant $H_{0}$ within the range $65 - 75$ km$\cdot$s$^{-1}$Mpc$^{-1}$ 
(in practical calculations, for example, for the analysis 
of the CMB anisotropy spectrum, these changes 
may be ignored at modern level of experimental data; these results are not 
presented here graphically). 
The values of $N_{e}/N_{H}$ calculated by varying the fraction of 
baryonic matter $\Omega_{b}$ are presented in Fig. \ref{omega_bar_var_ps} 
(top panel). $N_{e}/N_{H}$ is a monotonic function of $\Omega_{b}$. 
The HeII$\rightarrow$HeI recombination occurs at earlier epochs as 
$\Omega_{b}$ increases. Figure \ref{omega_bar_var_ps} (bottom panel) 
presents the relative difference between the value of $N_{e}/N_{H}$ 
calculated using our model and the value of $N_{e}/N_{H}$ calculated 
using the model by Wong et al.(2008) (i.e., in the Wong-Moss-Scott 
approximation).
We see from \ref{omega_bar_var_ps} that the results of the
calculations based on our model and the model by
Wong et al. (2008) agree with a relative accuracy
of at least $2.2\cdot 10^{-3}$ (in the number density of free
electrons) as the baryonic matter density varies in the
range $\Omega_b=0.02 - 0.06$.

\section{Conclusions}
We calculated the HeII$\rightarrow$HeI recombination kinetics
by including the effect of neutral hydrogen. We 
additionally took into account the partial redistribution
of HeI resonance photons in frequency upon their
scattering in the HeI $2^1P\rightarrow 1^1S$ line 
(Eqs. \ref{P_H_int3} - \ref{f_limit_final}). 

It is shown that the calculated relative numbers 
of free electrons $N_e/N_H$ could be satisfactorily reconciled 
with the results of recent studies of the HeII$\rightarrow$HeI 
recombination kinetics (Rubino-Martin et al. 2007; Wong et al. 2008) 
using the formulas derived here. 
The achieved accuracy is high enough for the
HeII$\rightarrow$HeI recombination model suggested here to
be used in analyzing the CMB anisotropy data from
future experiments (Planck and others). This accuracy
is also high enough to calculate the intensities,
frequencies, and profiles of the HeI recombination
lines formed during the cosmological HeII$\rightarrow$HeI recombination 
(Dubrovich and Stolyarov 1997; KhIV07; Rubino-Martin et al. 2007).

It should also be noted that using Eqs. (\ref{P_H_D_approx}), 
(\ref{approx_P_H}), 
(\ref{P_H_W}), (\ref{P_H_int3}), (\ref{P_H_R2} - \ref{f_limit_final}) 
derived here, along with Eq. (\ref{P_H_G}) derived by Grachev (1988), 
we can find not only the dependence of $P^{H}$ on 
$\gamma$, but also the dependences of $P^{H}$ on the Voigt parameter $a$ 
and the fraction of coherent scatterings $\lambda$ 
(and, accordingly, the fraction of incoherent scatterings $(1-\lambda)$, 
if the atomic transitions produced by electron collisions
may be neglected). This will be important in
further detailed theoretical studies of the HeII$\rightarrow$HeI
recombination.

\section*{Acknowledgments}
This work was supported by the Russian Foundation
for Basic Research (project no. 08-02-01246a) 
and the ``Leading Scientific Schools of Russia'' Program (NSh-2600.2008.2)
We wish to thank J.A. Rubino-Martin, J. Chluba, and R.A. Sunyaev for 
the results of calculations of the hydrogen-helium 
plasma ionization fraction provided for comparison.

\newpage
\section{References} 
1. M. M. Basko, Zh. Eksp. Teor. Fiz. 75, 1278 (1978) 
[Sov. Phys. JETP 48, 644 (1978)].
\\2. M. S. Burgin, Astron. Zh. 80, 771 (2003) [Astron. Rep. 47, 709 (2003)].
\\3. M.S.Burgin, V. L.Kauts, and N.N.Shakhvorostova, 
Pis'ma Astron. Zh. 32, 563 (2006) [Astron. Lett. 32, 507 (2006)].
\\4. J. Chluba and R. A. Sunyaev, Astron. Astrophys. 446, 39 (2006).
\\5. J. Chluba and R. A. Sunyaev, Astron. Astrophys. 475, 109 (2007).
\\6. J. Chluba and R. A. Sunyaev, Astron. Astrophys. 478, L27 (2008).
\\7. J. Chluba and R. A. Sunyaev, Astron. Astrophys. 480, 629 (2008).
\\8. N. N. Chugai, Astrofizika 26, 89 (1987).
\\9. V. K. Dubrovich, Pis'ma Astron. Zh. 1, 3 (1975) 
[Sov. Astron. Lett. 1, 1 (1975)].
\\10. V. K. Dubrovich and S. I. Grachev, Pis'ma Astron. Zh. 31, 403 (2005) 
[Astron. Lett. 31, 359 (2005)].
\\11. V. K. Dubrovich and V. A. Stolyarov, Pis'ma Astron.
Zh. 23, 643 (1997) [Astron. Lett. 23, 565 (1997)].
\\12. D. Galli and F. Palla, Planet. Space Sci. 50, 1197 (2002).
\\13. S. I. Grachev, Astrofizika 28, 205 (1988).
\\14. S. I. Grachev and V. K. Dubrovich, Astrofizika 34, 249 (1991).
\\15. S. I. Grachev and V. K. Dubrovich, Astronomy Letters, 34, 439 (2008). 
\\16. J. P. Harrington, Mon. Not. R. Astron. Soc. 162, 43 (1973). 
\\17. C.M. Hirata, Physical Review D, 78, id. 023001 (2008). 
\\18. C. M. Hirata and E. R. Switzer, Phys. Rev. D 77, 083007 (2008).
\\19. V. V. Ivanov, Radiative Transfer and the Spectra of 
Celestial Bodies (Nauka, Moscow, 1969) [in Russian].
\\20. B. J. T. Jones and R. F. G. Wyse, Astron. Astrophys. 149, 144 (1985).
\\21. E. E. Kholupenko and A. V. Ivanchik, Pis'ma Astron. Zh. 
32, 12, 883 (2006) [Astron. Lett. 32, 795 (2006)].
\\22. E. E. Kholupenko, A. V. Ivanchik, and D. A. Varshalovich, 
Mon. Not. R. Astron. Soc. Lett. 378, L39 (2007); astro-ph/0703438.
\\23. K. R. Lang, Astrophysical Formulae (Mir, Moscow, 1978; Springer-Verlag, 
New York, 2002). 
\\24. P. K. Leung, C. W. Chan, and M.-C. Chu, Mon. Not. R. Astron. Soc. 
349, 632 (2004).
\\25. T. Matsuda, H. Sato, and H. Takeda, Progr. Theor. Phys. 42, 219 (1969).
\\26. D. I. Nagirner, Lectures on the Theory of Radiative Transfer 
(St. Petersburg State Univ., St. Petersburg, 2001) [in Russian].
\\27. B. Novosyadlyj, Mon. Not. R. Astron. Soc. 370, 1771 (2006).
\\28. P. J. Peebles, Astrophys. J. 142, 1317 (1965).
\\29. P. J. Peebles, Astrophys. J. 153, 1 (1968).
\\30. J. A. Rubino-Martin, J. Chluba, and R. A. Sunyaev, arXiv:0711.0594 (2007).
\\31. G. B. Rybicki and I. P. dell' Antonio, ASP Conf. Ser. 51, 548 (1993).
\\32. S. Seager, D. Sasselov, and D. Scott, Astrophys. J. Lett. 523, L1 (1999).
\\33. S. Seager, D. Sasselov, and D. Scott, Astrophys. J. Suppl. Ser. 128, 407 (2000).
\\34. P. C. Stancil, A. Loeb, M. Zaldarriaga, et al., Astrophys. J. 580, 29 (2002).
\\35. R. A. Sunyaev and J. Chluba, arXiv:0802.0772 (2008).
\\36. E. R. Switzer and C. M. Hirata, Phys. Dev. D. 77, 083006 (2008).
\\37. D. A. Varshalovich and R. A. Sunyaev, Astrofizika 4, 359 (1968).
\\38. W. Y. Wong and D. Scott, Mon. Not. R. Astron. Soc. 375, 1441 (2007).
\\39. W. Y. Wong, A. Moss, and D. Scott, Mon. Not. R. Astron. Soc. 386, 1023 (2008).
\\40. Ya. B. Zeldovich, V. G. Kurt, and R. A. Sunyaev, 
Zh. Eksp. Teor. Fiz. 55, 278 (1968) [Sov. Phys. JETP 28, 146 (1968)].

\newpage

\begin{figure*}
\centering
\includegraphics[height=12cm, width=12cm]{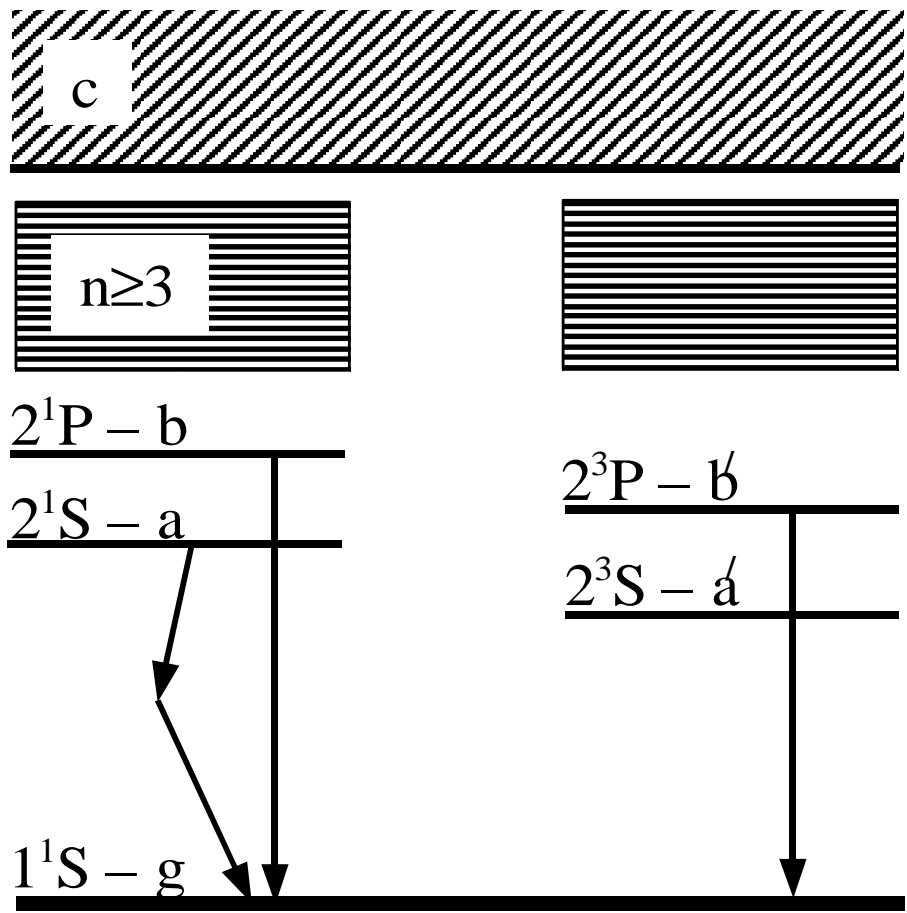}
\caption{Model energy level diagram for the HeI atom used here.}
\label{level_scheme}
\end{figure*}

\newpage
\begin{figure*}
\centering
\includegraphics[width=16cm, bb=15 6 550 540]{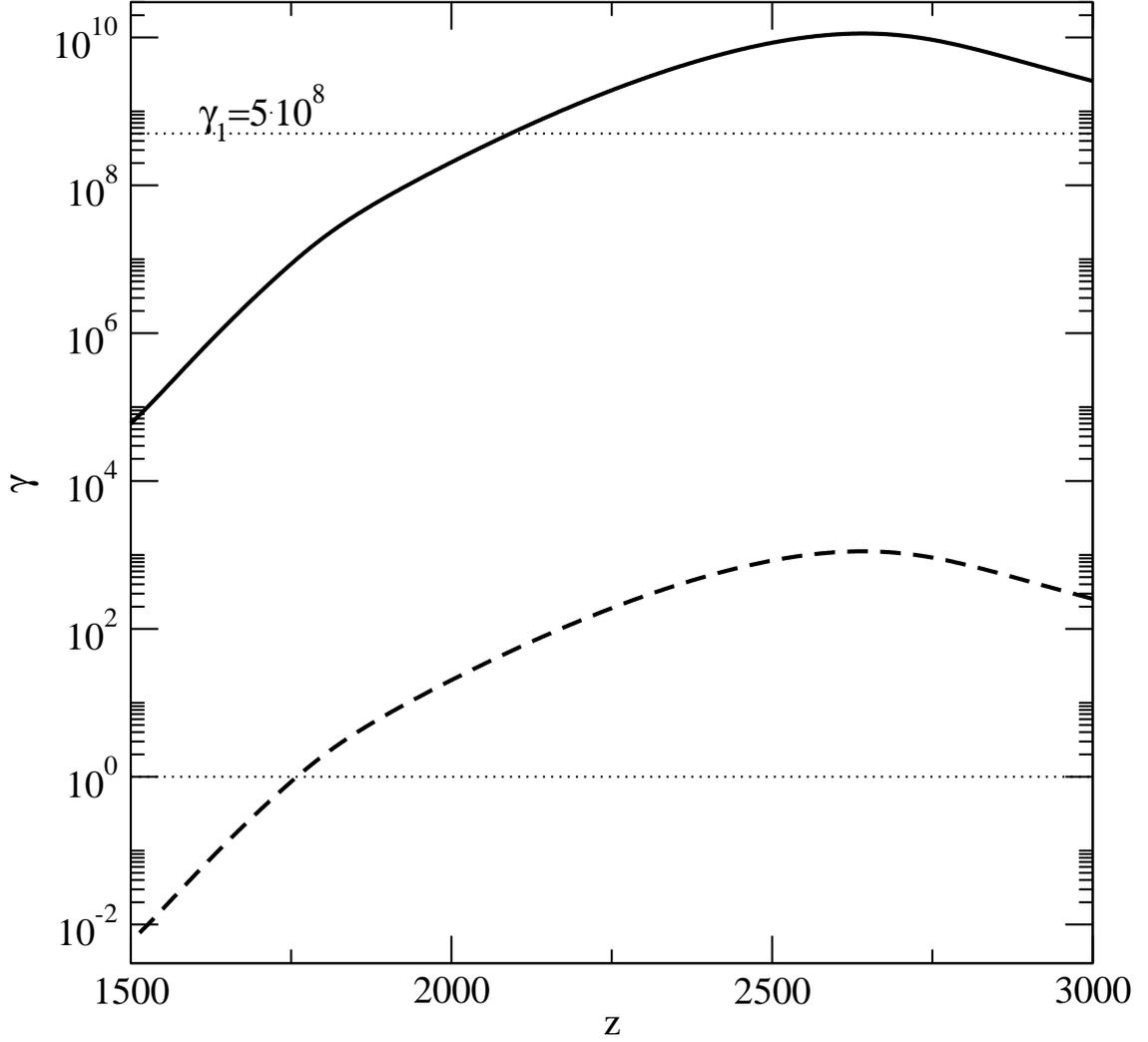}
\caption{Ratio of the HeI and HI absorption coefficients
at the central frequency of the $f\rightarrow g$ line ($\gamma$) versus
redshift $z$ for the HeI $2^1P\rightarrow 1^1S$ (solid curve) and 
HeI $2^3P\rightarrow 1^1S$ (dashed curve) transitions. The dotted
straight lines indicate the unit and $\gamma_1=5\cdot 10^8$ levels. 
The value of $\gamma_1=5\cdot 10^8$ is explained in the text.}
\label{gammas_ps}
\end{figure*}

\newpage
\begin{figure*}[t]
\centering
\includegraphics[width=16cm, bb=15 6 550 540]{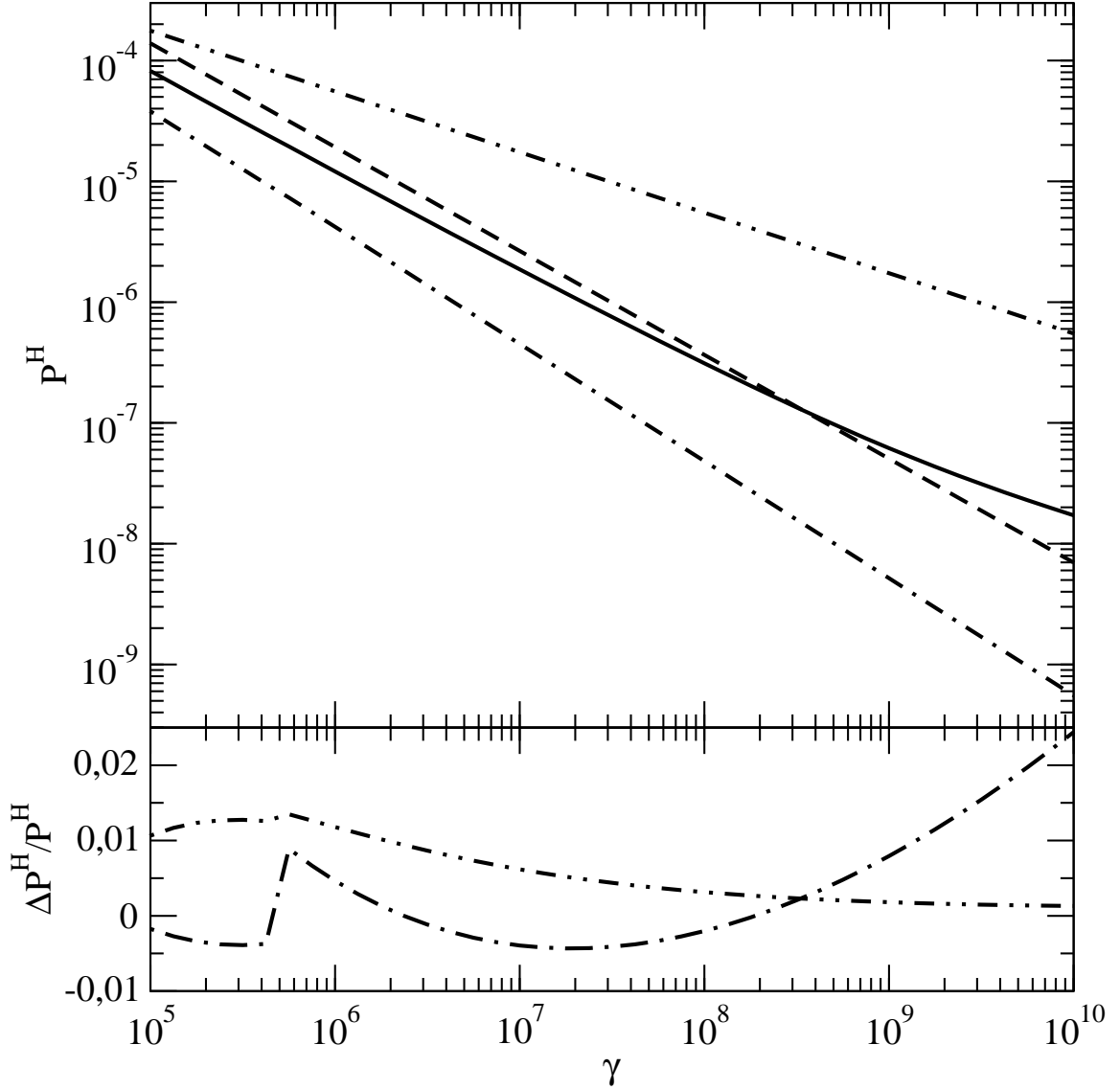}
\caption{Top panel: Destruction probability of a HeI $2^1P\rightarrow 1^1S$ 
resonance photon as it interacts with neutral hydrogen $P^H$ versus $\gamma$ 
for various models of absorption and redistribution in frequency in the HeI 
$2^1P\rightarrow 1^1S$ line: 
the dash-dotted curve corresponds to the use of the Doppler
absorption profile, the solid curve corresponds to a partial
redistribution in frequency (Eq. (\ref{P_H_int3})), 
the dashed curve represents the Wong-Moss-Scott approximation, and 
the dashed curve with two dots corresponds to the use
of the Voigt absorption profile for a complete redistribution
in frequency. The Voigt parameter is $a=1.7\cdot 10^{-3}$, 
and the single-scattering albedo is defined by the relation 
$(1-\lambda)=2.5\cdot 10^{-3}$. 
Bottom panel: The relative error in $P^H$ when calculated from the 
approximate formulas (\ref{approx_P_H}) (the
dashed curve with two dots corresponding to the use of
the Voigt absorption profile for a complete redistribution
in frequency) and (\ref{P_H_D_approx}) (the dash-dotted curve corresponding
to the use of the Doppler absorption profile) compared to the numerical 
calculation based on the exact formula (\ref{P_H_int2}).}
\label{para_P_gamma_ps}
\end{figure*}

\newpage
\begin{figure*}[t]
\centering
\includegraphics[width=16cm, bb=15 6 550 540]{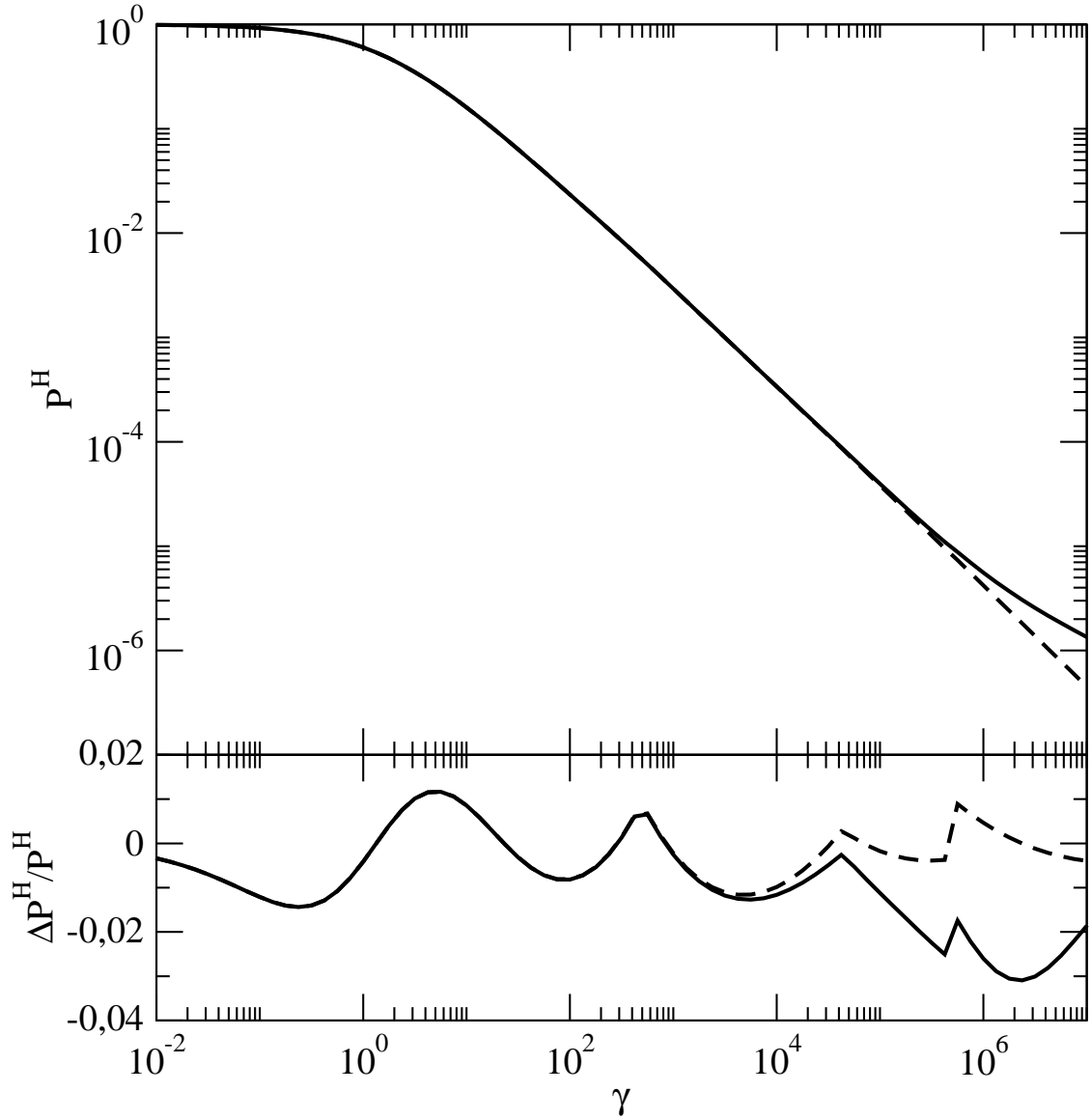}
\caption{Top panel: Destruction probability of a HeI $2^3P\rightarrow 1^1S$ 
resonance photon as it interacts with neutral hydrogen $P^H$ versus $\gamma$ 
for various absorption profiles in the HeI $2^3P\rightarrow 1^1S$ line: 
the dashed curve corresponds to the
use of the Doppler absorption profile and the solid curve
corresponds to the use of the Voigt absorption profile
for a complete redistribution in frequency. The Voigt parameter
is $a=10^{-5}$ and the single-scattering albedo is
approximately equal to zero $\lambda \simeq 0$
(typical values for the HeI $2^3P\rightarrow 1^1S$ transition at the 
HeII$\rightarrow$HeI recombination epoch). 
Bottom panel: The relative error in $P^H$ when
calculated from the approximate formulas (\ref{approx_P_H}) (the solid
curve corresponding to the use of the Voigt absorption
profile for a complete redistribution in frequency) and (\ref{P_H_D_approx}) 
(the dashed curve corresponding to the use of the Doppler
absorption profile) compared to the numerical calculation
based on the exact formula (\ref{P_H_int2}).}
\label{orto_P_gamma_ps}
\end{figure*}

\newpage
\begin{figure*}
\centering
\includegraphics[width=16cm, bb=15 6 550 540]{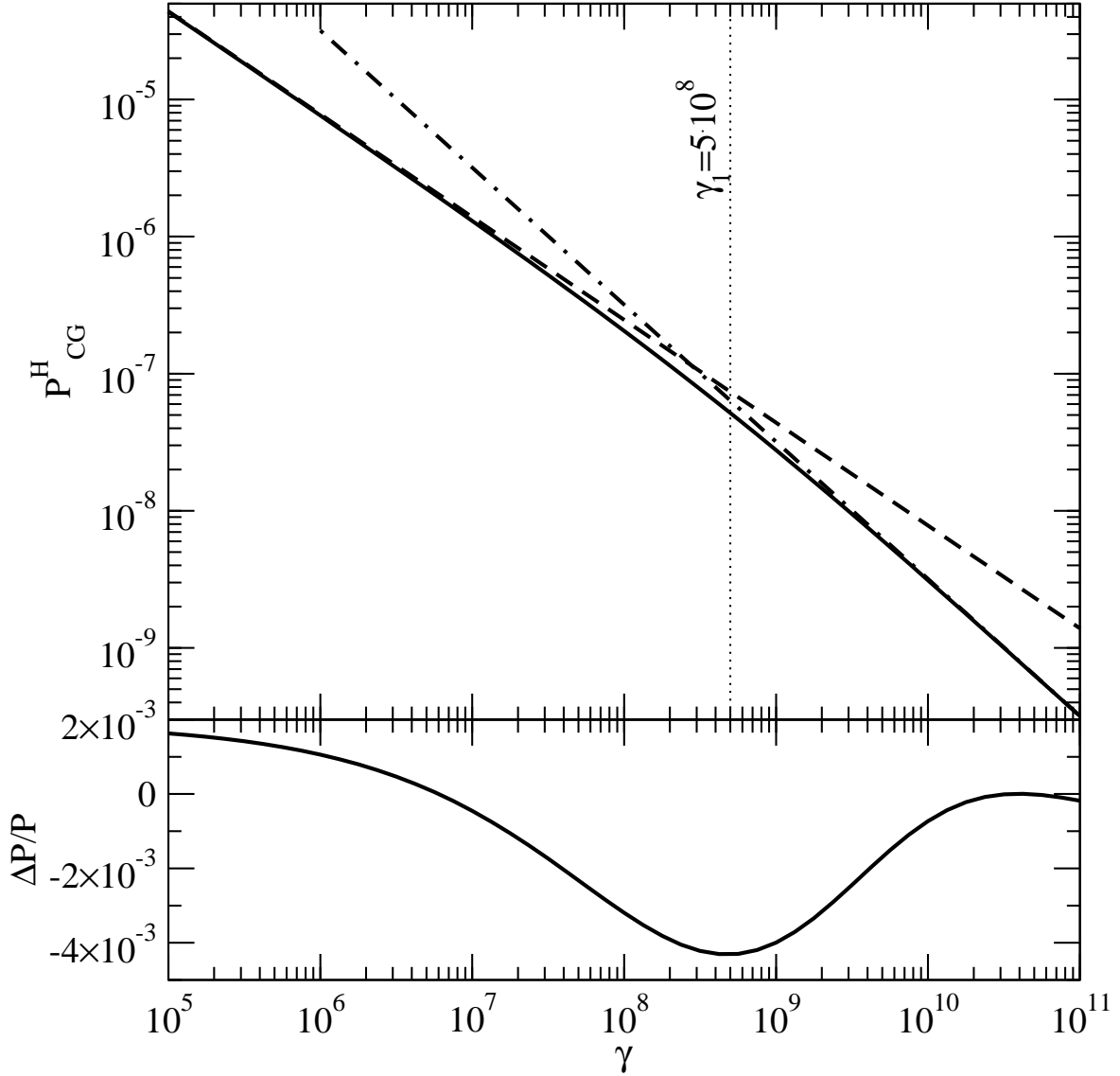}
\caption{Top panel: Destruction probability of a HeI 
$2^1P\rightarrow 1^1S$ resonance photon as it interacts with neutral hydrogen 
$P^{H}_{G}$ (\ref{P_H_G}) versus $\gamma$ (solid curve), asymptotics 
(\ref{P_H_G_low}) corresponding to $\gamma\ll\gamma_1$ (dashed curve) 
(Chugai 1987; see the text), and asymptotics (\ref{P_H_G_high}) 
corresponding to $\gamma\gg\gamma_1$ (dash-dotted curve). 
The Voigt parameters is $a=1.7\cdot 10^{-3}$ and the single-scattering 
albedo is defined by the relation $(1-\lambda)=2.5\cdot 10^{-3}$ 
(typical values for the HeI $2^1P\rightarrow 1^1S$ transition at 
the HeII$\rightarrow$HeI recombination epoch). 
Bottom panel: The relative error in $P^{H}_{G}$ approximately calculated
from Eqs. (\ref{Gamma_approx}) and (\ref{F_approx}) 
compared to the calculations based on Eq. (\ref{P_H_G}).}
\label{P_CG_ps}
\end{figure*}

\newpage
\begin{figure*}[t]
\centering
\includegraphics[width=12cm, bb=15 6 550 540]{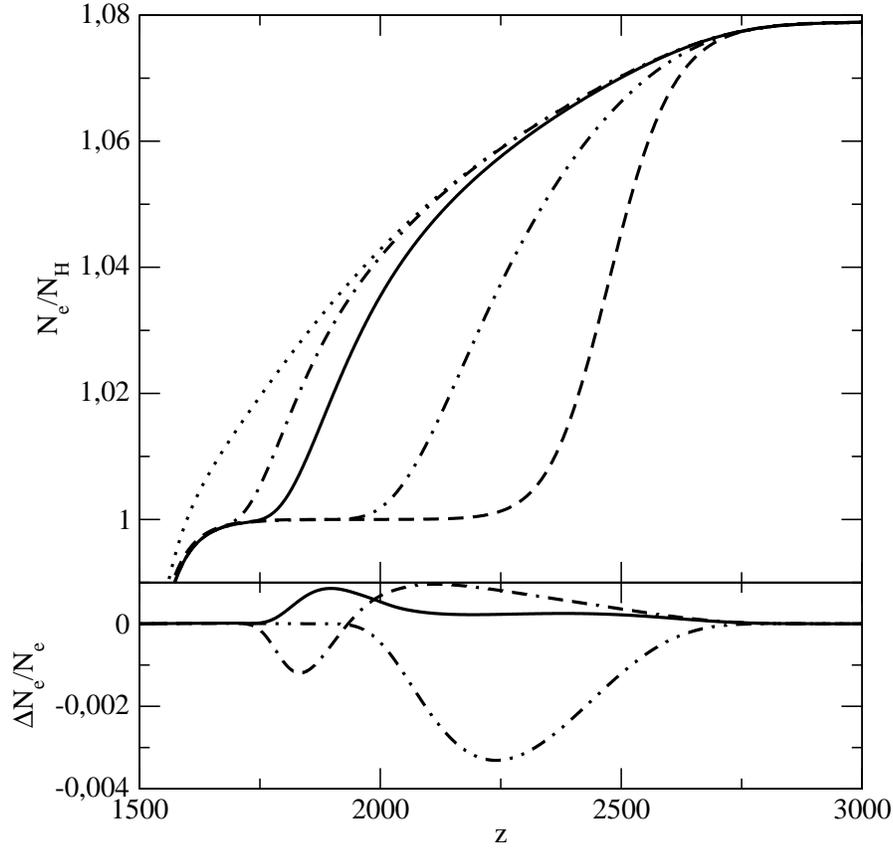}
\caption{Top panel: Relative number of free electrons $N_{e}/N_{H}$ 
versus redshift $z$: the dashed curve represents the equilibrium
recombination, the dotted curve represents the recombination
according to the model without any absorption of
HeI resonance photons by neutral hydrogen; the dash-dotted curve 
represents the recombination according to 
the model by KhIV07; the solid curve
represents the recombination according to this work 
(Eqs. (\ref{approx_P_H}) and (\ref{P_H_int3}) describe the 
HeI $2^3P\rightarrow 1^1S$ and HeI $2^1P\rightarrow 1^1S$ transition 
probabilities, respectively); 
the dashed curve with two dots represents the recombination 
calculated using the approximation of a complete 
redistribution in frequency in HeI resonance lines 
(Eq. (\ref{approx_P_H}) describes the probability of both 
HeI $2^3P\rightarrow 1^1S$ and HeI $2^1P\rightarrow 1^1S$ transitions). 
All curves were calculated for the cosmological parameters adopted in
Rubino-Martin et al. (2007).
Bottom panel: The relative difference
in the fraction of free electrons for different recombination
models: the solid curve indicates the difference between
our results and the results by Rubino-Martin et al. (2007)
for the models with a partial redistribution in frequency;
the dashed curve with two dots indicates the difference
between our results and the results of Rubino-Martin
et al. (2007) for the models with a complete redistribution
in frequency; the dash-dash-dot curve indicates
the difference between the results based on the model
by Wong et al. (2008) and those by Rubino-Martin
et al. (2007) for the model with a partial redistribution in
frequency.}
\label{var_model_ps}
\end{figure*}

\newpage
\begin{figure*}
\centering
\includegraphics[width=16cm, bb=15 6 550 540]{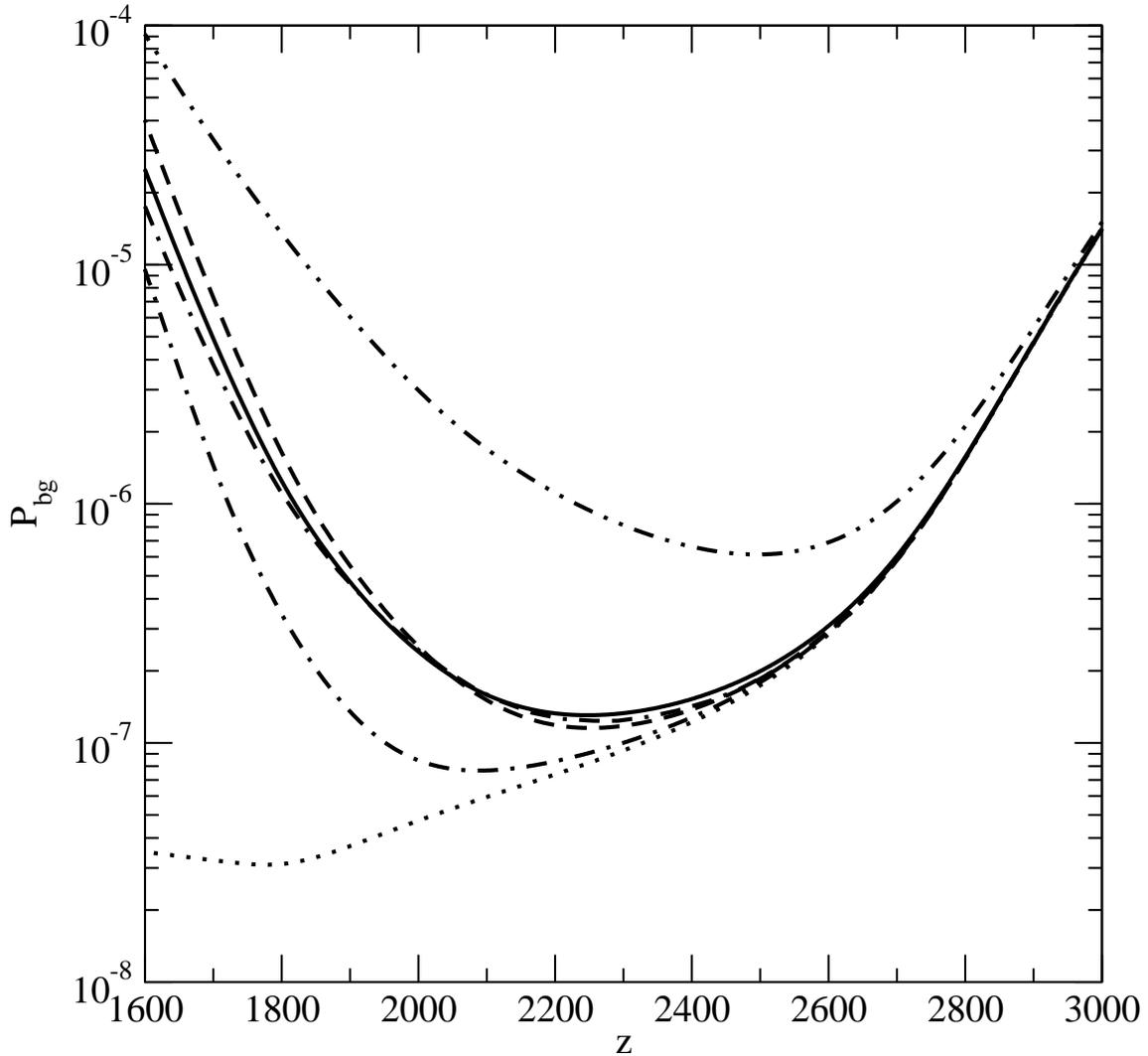}
\caption{Probability of the uncompensated HeI $2^1P\rightarrow 1^1S$ 
transitions $P_{bg}$ versus redshift $z$ for various models
of absorption and redistribution in frequency in HeI 
$2^1P\rightarrow 1^1S$ line: 
the dash-dotted curve corresponds to
the use of the Doppler absorption profile, the solid curve
represents a partial redistribution in frequency (this paper),
the dash-dash-dot curve represents a partial redistribution
in frequency (Rubino-Martin et al. 2007), the
dashed curve represents the Wong-Moss-Scott approximation,
the dashed curve with two dots corresponds to
the use of the Voigt absorption profile for a complete redistribution
in frequency, and the dotted curve represents
the Sobolev probability $P^{S}_{bg}$ (calculated from Eq. 
(\ref{Sobolev_probability})). 
The curves were calculated for the cosmological parameters
adopted in Rubino-Martin et al. (2007).}
\label{para_P_z_ps}
\end{figure*}

\newpage
\begin{figure*}
\centering
\includegraphics[width=16cm, bb=15 6 550 540]{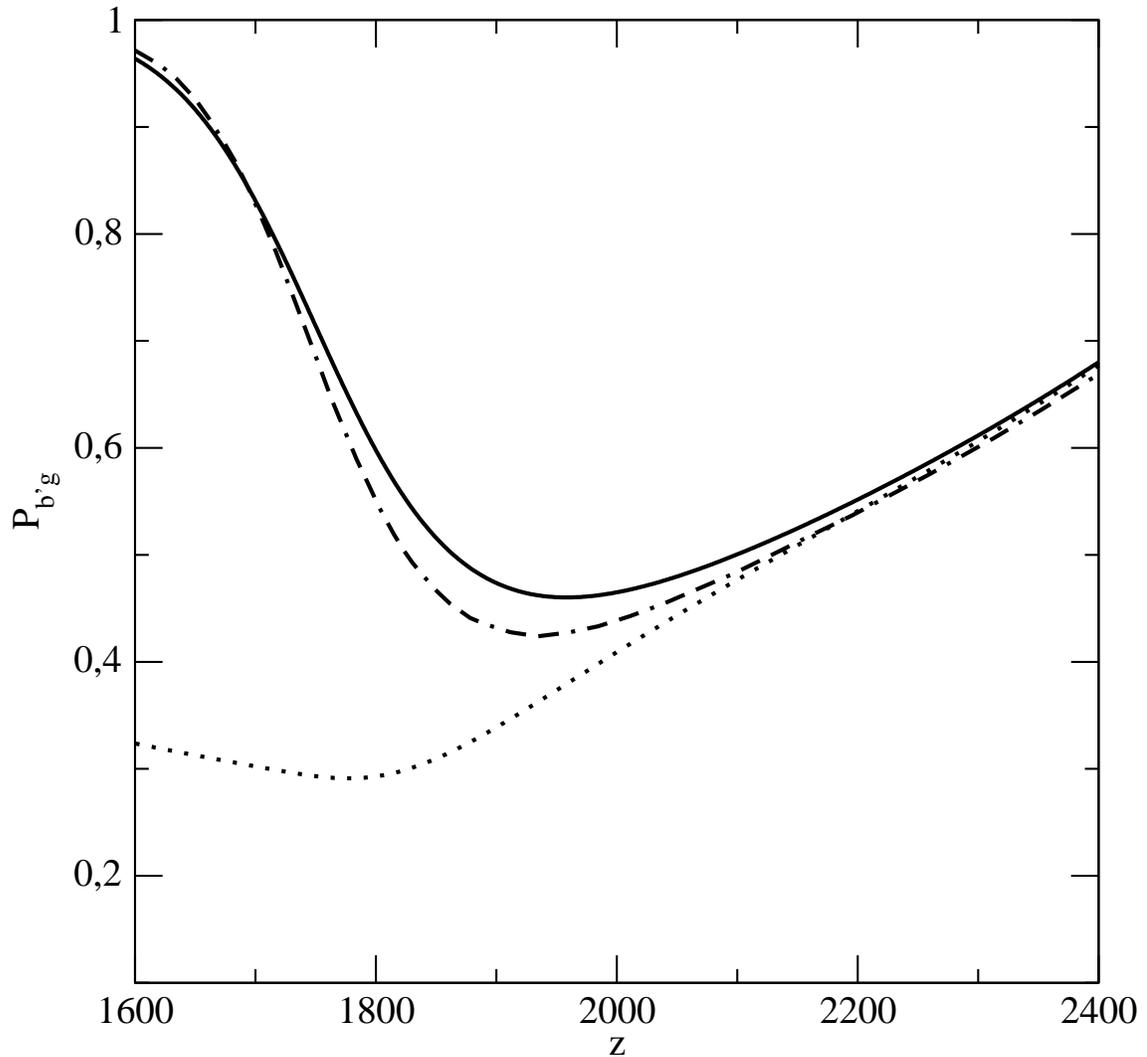}
\caption{Probability of the uncompensated HeI $2^3P\rightarrow 1^1S$ 
transitions $P_{b'g}$ versus redshift $z$: 
the solid curve represents our result (Eq. (\ref{approx_P_H}) 
was used for the calculations, which corresponds to the use of 
the Voigt absorption
profile for a complete redistribution in frequency), the
dash-dash-dot curve represents the result by Rubino-Martin et al. (2007), 
and the dotted curve represents the Sobolev probability 
$P^{S}_{b'g}$ (calculated from Eq. (\ref{Sobolev_probability})). 
The curves were calculated for the cosmological parameters 
adopted in Rubino-Martin et al. (2007).} 
\label{orto_P_z_ps}
\end{figure*}

\newpage
\begin{figure*}
\centering
\includegraphics[width=16cm, bb=15 6 550 540]{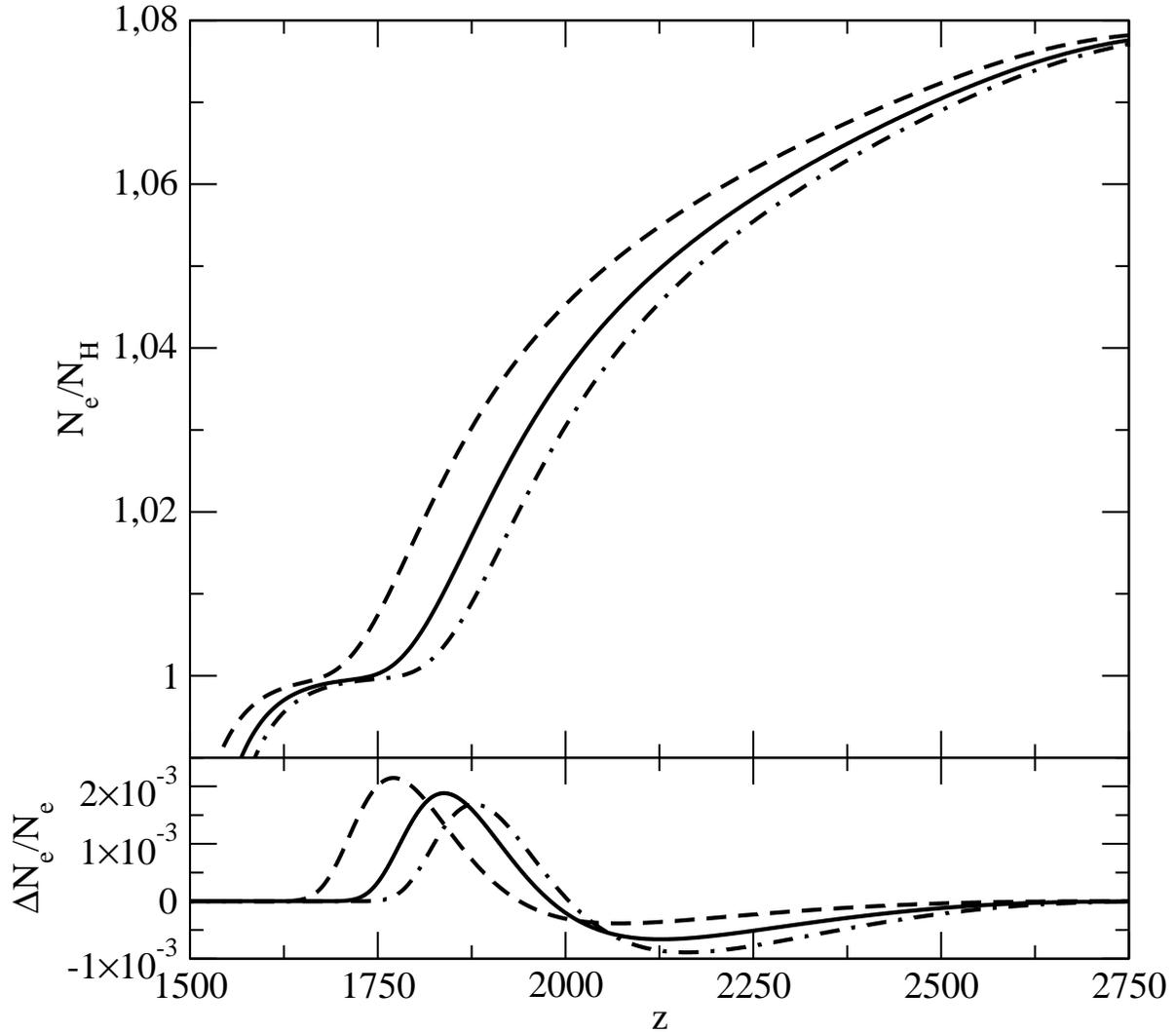}
\caption{Top panel: Relative number of free electrons $N_{e}/N_{H}$ 
versus redshift $z$ for various baryonic matter densities:
the dashed, solid, and dash-dotted curves correspond to 
$\Omega_b=0.02$, $\Omega_b=0.04$, and $\Omega_b=0.06$, respectively. 
The remaining parameters of the cosmological model
are given in Table \ref{cosm_par}. 
Bottom panel: The relative difference in the 
number density of free electrons between the calculations 
using our model and those using the model by Wong et al. (2008) 
(the Wong-Moss-Scott approximation):
the dashed, solid, and dash-dotted curves correspond to  
$\Omega_b=0.02$, $\Omega_b=0.04$, and $\Omega_b=0.06$, respectively.}
\label{omega_bar_var_ps}
\end{figure*}

\end{document}